\documentclass[aps,pra,nofootinbib,preprint]{revtex4-1}

\usepackage{graphicx}

\usepackage{subfig}
\usepackage{amsmath}	
\usepackage{amssymb}    

\newcommand{\be}{\begin{eqnarray}}
 \newcommand{\ee}{\end{eqnarray}}

\newcommand{\R}{\mathbb{R}}

\begin{document}



\title{Relativistic Stark resonances in a simple exactly soluble model for a diatomic molecule}

\author{Fran\c{c}ois Fillion-Gourdeau}
\email{filliong@CRM.UMontreal.ca}
\affiliation{Centre de Recherches Math\'{e}matiques, Universit\'{e} de Montr\'{e}al, Montr\'{e}al, Canada, H3T 1J4}
\altaffiliation[Also at ]{School of Mathematics and Statistics, Carleton University, Ottawa, Canada, K1S 5B6; and \\Fields Institute, University of Toronto, Toronto, Canada, M5T 3J1}

\author{Emmanuel Lorin}
\email{elorin@math.carleton.ca}
\affiliation{School of Mathematics and Statistics, Carleton University, Ottawa, Canada, K1S 5B6}
\altaffiliation[Also at ]{Centre de Recherches Math\'{e}matiques, Universit\'{e} de Montr\'{e}al, Montr\'{e}al, Canada, H3T 1J4}

\author{Andr\'{e} D. Bandrauk}
\email{Andre.Dieter.Bandrauk@USherbrooke.ca}
\affiliation{Laboratoire de chimie th\'{e}orique, Facult\'{e} des Sciences, Universit\'{e} de Sherbrooke, Sherbrooke, Canada, J1K 2R1}
\altaffiliation[Also at ]{Centre de Recherches Math\'{e}matiques, Universit\'{e} de Montr\'{e}al, Montr\'{e}al, Canada, H3T 1J4}

\date{\today}

\begin{abstract}
A simple 1-D relativistic model for a diatomic molecule with a double point interaction potential is solved exactly in a constant electric field. The Weyl-Titchmarsh-Kodaira method is used to evaluate the spectral density function, allowing the correct normalization of continuum states. The boundary conditions at the potential wells are evaluated using Colombeau's generalized function theory along with charge conjugation invariance and general properties of self-adjoint extensions for point-like interactions. The resulting spectral density function exhibits resonances for quasibound states which move in the complex energy plane as the model parameters are varied. It is observed that for a monotonically increasing interatomic distance, the ground state resonance can either go deeper into the negative continuum or can give rise to a sequence of avoided crossings, depending on the strength of the potential wells. For sufficiently low electric field strength or small interatomic distance, the behavior of resonances is qualitatively similar to non-relativistic results. 
\end{abstract}


\maketitle


\pagestyle{plain}


\section{Introduction}

The interaction of matter with very intense laser field has been a very active field of research in the last few decades and led to the discovery of very interesting phenomena such as above-threshold ionization, high harmonic generation and others \cite{Ehlotzky199863,RevModPhys.72.545}. These studies were motivated mostly by the advent of new lasers reaching high intensity levels. In that regime, the traditional theoretical tools such as perturbation theory are not applicable and thus, many other avenues were explored to study these systems mathematically \cite{Ehlotzky199863,RevModPhys.72.545}. More recently, the development in laser technologies have allowed to reach unprecedented intensity levels ($10^{20}\; \mbox{W}/\mbox{cm}^{2}$ and higher \cite{RevModPhys.78.309}) providing an opportunity to study a new range of nonperturbative parameters \cite{Salamin200641}. The interaction of these lasers with matter, such as electrons, atoms or molecules, allows to probe relativistic effects: the ponderomotive energy of an electron in that regime has an order of magnitude close to the rest mass energy \cite{Salamin200641,0034-4885-72-4-046401}. In this setting, the Dirac equation should be used to give a consistent description of these phenomena instead of the non-relativistic Schr\"odinger equation. 

From a mathematical point of view, finding a solution to the Dirac equation is a very challenging task because of its intricate matrix structure. For this reason, exact solutions can be found only for a few special cases describing highly symmetric systems \cite{Greiner:1987,Greiner:1985,bagrov1990exact}. Approximate solutions however can be determined through semi-classical techniques and in some cases, realistic systems can be analyzed by using these analytical methods \cite{PhysRevLett.89.193001}.  Another approach is to use numerical methods; this is the subject of numerous studies for both the time-dependent \cite{Salamin200641} and time-independent cases \cite{esteban2007,Desclaux2003453}. However, these solutions often require an important amount of computational resources and the numerical methods require special care to circumvent numerical artifacts. Moreover, these calculations are often very complicated, albeit being more realistic, and the correct physical interpretation is often harder to reach. For these reasons, it may be helpful to consider simpler models to understand the basic physical features of a given system. In this article, we adhere to this philosophy: we are considering a simplified relativistic model for a diatomic molecule in a quasi-static electric field. More precisely, the molecule is modeled by two attracting point-like potentials (double delta function potential) and is subjected to a constant electric field. This approximates in a crude way the short-range Coulomb potential of the nuclei and its interaction with a slowly varying electromagnetic laser field. To simplify the analysis further, we consider a one-dimensional model for which analytical solutions can be determined. This last approximation can also be justified in the strong field limit where ionization along the electric field is the more important mechanism in which case a ``real'' linear polyatomic molecule behaves almost like a 1-D system \cite{PhysRevLett.90.233003,PhysRevLett.98.013001}.  

This kind of simplified models using point-like interaction has been studied extensively both relativistically \cite{0305-4470-22-10-001,PhysRevA.32.1208,PhysRevA.24.1194,springerlink:10.1007/BF00401163,springerlink:10.1007/BF01228101,PSSB:PSSB196} and non-relativistically \cite{0305-4470-30-11-021,albeverio2005solvable} for scattering and bound states problems when the electric field is absent. In these cases, an analytical solution can be calculated easily. In the presence of a constant homogeneous electric field, the mathematical problem is more challenging because the differential operator becomes singular on the domain limits (at $x=\pm \infty$), which complicates the boundary condition prescription. Also, the continuum spectrum $\sigma_{c}$ of the differential operator changes dramatically: there is a continuum on the whole range of energy ($\sigma_{c} = \R$) while bound states become resonances for sufficiently small field strength (a resonance corresponds to a pole of the resolvent operator on the ``unphysical'' Riemann sheet \cite{reed1972methods}). This is to be contrasted with the zero field case in which the continuum spectrum is $\sigma_{c} = (-\infty,-mc^{2}] \cup [mc^{2},\infty) $, the bound states in this case are points in $(-mc^{2},mc^{2})$. To cope with these difficulties, analytical methods have been developed and applied to these simple models. For instance, Green's function methods were used in \cite{PhysRevB.42.7630,0305-4470-36-48-009,0305-4470-36-8-311,0305-4470-36-8-311} to compute the Stark energy shift and decay rates of quasibound states. The Weyl-Titchmarsh-Kodaira (WTK) method \cite{1962eeaw.bookT,kodaira1949}, which allows to compute the spectral density of second-order singular operators, was used to analyze many one-dimensional models \cite{dean:540,10.1063/1.526733} and to develop numerical methods for arbitrary 1-D potentials \cite{PhysRevLett.81.2663}. The same kind of approach was used in \cite{PhysRevA.10.1494} to evaluate the non-perturbative Stark effect in hydrogen-like atoms. Finally, the 1-D non-relativistic diatomic molecule with point-like interactions in a constant field has been investigated in \cite{PhysRevA.69.053409} using this WTK technique. In their analysis, the authors were able to compute analytically the spectral density and showed results in very good qualitative agreement with real diatomic molecules. In our study, a similar approach is used to study the latter in the relativistic regime using the extension of the WTK method to the Dirac equation \cite{Titchmarsh01011961,Titchmarsh01011961}. This methodology has been used in \cite{PhysRevLett.101.190401,Riccardo2011,1402-4896-75-5-009,springerlink:10.1140/epjc/s10052-009-0999-x} to study various potentials, while the atomic case with a single delta function potential was treated in \cite{GonzalezDiaz2006202}.

An important advantage of working with point-like potentials is that they can be characterized by boundary conditions on the Dirac potential well positions. This procedure in the non-relativistic case is well-known and a rigorous discussion of this subject can be found in \cite{albeverio2005solvable}. For the Dirac equation, this has been the subject of debates in the last few decades because of ambiguous results: different methods yielded different boundary conditions \cite{PhysRevA.24.1194}. This ambiguity was related to the fact that the wave function has a jump discontinuity and is multiplied by a delta function, resulting in an ill-defined product of distributions \cite{calkin:737,springerlink:10.1007/BF00750071}. This was also studied rigorously by looking at the self-adjoint extensions of the Dirac operator and it was shown that there exists a four-parameter family of self-adjoint operators corresponding to point interactions \cite{springerlink:10.1007/BF00401163,springerlink:10.1007/BF00750071,Benvegnu1994,Hughes1997425,springerlink:10.1007/BF00397060,springerlink:10.1007/BF00420304}. In this work, we make use of certain results from the algebra of Colombeau's generalized function to deal with products of distributions \cite{Colombeau198396,Colombeau1990,colombeau:315,Colombeau:1985}. By combining these ideas with general properties of self-adjoint extensions \cite{Benvegnu1994} and charge conjugation invariance, we show that it is possible to select a specific boundary condition which corresponds to an electrostatic Dirac potential wells.

This article is separated as follows. In Section \ref{sec:wtk}, the WTK method is shortly presented. This discussion focuses on general results which allow to comprehend the general ideas and to understand how to apply the methods in our case. In Section \ref{sec:free}, the free case (with no electric field nor point-like potential) is considered to show an implementation of the WTK method and an important qualitative difference with the non-relativistic case, i.e. the appearance of the negative energy continuum. The diatomic molecule without electric field is considered in Section \ref{sec:free_delta} as a consistency check for the full case, which is treated in Section \ref{sec:free_delta}. The calculation of boundary conditions using Colombeau's theory is relegated to Appendix \ref{sec:app_bound}, along with a few comments on the Dal Maso-Le Floch-Murat definition of products of distribution as a bounded Borel measure. Note that throughout this work, we are working in units where $c=m=\hbar=1$, so that the Compton radius is unity ($m$ and $c$ are kept explicitly in all equations to easily switch to atomic or natural units).

\section{General theory of eigenfunction expansion for the Dirac equation}
\label{sec:wtk}

In this section, a brief review of the WTK method is presented. This discussion focuses on main results and we refer the interested reader to the extensive mathematical literature on the subject for more details \cite{1962eeaw.bookT,nla.cat-vn2169453,fulling1989aspects}.

To begin, let us consider an equation defined on a domain $x \in \mathbb{R}$ of the form
\begin{eqnarray}
 H \psi(x) = E \psi(x)
\end{eqnarray}
where $H$ is the Dirac operator, $E$ is the eigenenergy and $\psi \in L^{2}(\mathbb{R}) \otimes \mathbb{C}^{2}$ is the wave function (bi-spinor). The main objective here is to expand the wave function over an eigenfunction $(\psi_{E})_{E}$ basis. Assuming that $H$ is a self-adjoint linear operator on $\mathcal{D}(H)$ defined in a Hilbert space, the spectral theorem gives the spectral decomposition:
\begin{eqnarray}
\label{eq:spec_th}
 H = \int_{\sigma(H)} E dP_{E} 
\end{eqnarray}
where $\sigma(H)$ is the spectrum of operator $H$ and $P_{E}$ is a projection operator onto the eigenspace spanned by the eigenvectors $\psi_{E}$. Written in this abstract form, the spectral theorem is of limited utility for practical applications: its mathematical properties can obviously be studied but it says little concerning the explicit expression of $P_{E}$. 

The WTK method provides an explicit realization of the spectral theorem for the Dirac operator in 1-D: it allows, from the knowledge of a solution of the equation, to construct a generalized eigenfunction expansion, in the form of (\ref{eq:spec_th}) \cite{10.1137/0513063}. At the same time, it yields the spectral density which has a very important physical meaning since the latter exhibits the main features of the spectrum (bound states, resonances and continuum part). Finally, the WTK method can be used to treat ``singular'' problems where the operator $H$ has a singularity on one or both domain boundaries. This case is relevant when the system is immersed in a static electric field, as considered in this study, and where we have $ \lim_{x \rightarrow \pm} V(x) = \mp \infty $ for the potential $V$. The main result of the WTK analysis is that any smooth function $f$ can be expanded as
\begin{eqnarray}
\label{eq:eigen_exp}
 f(x) = \int_{-\infty}^{\infty} \sum_{i,j=0}^{1} \gamma_{i}(x,E) \rho_{ij}(E) \tilde{f}_{j}(E) dE
\end{eqnarray} 
where $\rho_{ij}(E)$ are the components of the matrix-valued spectral density, $\gamma_{0}(x,E) := u^{\pm}(x)$, $\gamma_{1}(x,E) := v^{\pm}(x)$ (the functions $u$ and $v$ will be defined later, see (\ref{eq:init_cond_u}) and (\ref{eq:init_cond_v})) and 
\begin{eqnarray}
 \tilde{f}_{j}(E) = \int_{-\infty}^{\infty} \gamma^{\dagger}_{j}(x,E) f(x) dx
\end{eqnarray}
are Fourier-like coefficients. Note here that by applying the Hamiltonian $H$ to $f$, one obtains an expression for the projection operator $P_{E}$ in terms of the spectral density. The latter can be specified explicitly because they can be related to the classical solution. This procedure will be discussed further in the following.

The WTK method was originally developed to extend the Sturm-Liouville problem to infinite domain with singular boundary conditions, but it can be carried to the relativistic case and more generally, for the solution of any first order systems of equation of the form \cite{Titchmarsh01011961}
\begin{eqnarray}
\label{eq:gen_form}
p(x) \psi_{1}(x) + q(x) \psi_2(x) + \partial_x \psi_2(x) = E \psi_1(x) , \\
\label{eq:gen_form2}
q(x) \psi_{1}(x) + r(x) \psi_2(x) - \partial_x \psi_1(x) = E \psi_2(x) , 
\end{eqnarray}
for all $x \in \mathbb{R}$, where $p,r,q$ are real functions which can have singularities on the domain boundaries. The 1-D Dirac equation is a special case of the last equation. More specifically, we first consider the case where $x^{-} < x < x^{+}$ (we will let $x^{-} \rightarrow -\infty$ and $x^{+} \rightarrow \infty$ later) . The Dirac equation is given by
\begin{equation}
 E\psi(x) = \left[ -ic \alpha \partial_{x} + \beta mc^{2} + V(x)  \right] \psi(x), \;\; x \in [x^{-},x^{+}]
\end{equation}
where in 1-D, $\psi$ is a bi-spinor, $V$ is the scalar potential, $m$ is the fermion mass and $E$ is the energy. It is convenient to work in a representation where the Dirac matrices are given by
\begin{eqnarray}
 \alpha = -\sigma_{y} \;\; \mbox{and} \;\; \beta = -\sigma_z
\end{eqnarray}
which obey the appropriate anticommutation relations\footnote{This can be easily obtained from the usual Dirac representation by using a change of representation transformation defined as
\begin{eqnarray}
 U = \frac{1}{\sqrt{2}} (\sigma_{x} - \sigma_{y}) .
\end{eqnarray}
such that $\alpha_{\rm new} = U\alpha_{\rm dirac} U^{\dagger}$, $\beta_{\rm new} = U\beta_{\rm dirac} U^{\dagger}$ and $UU^{\dagger} = \mathbb{I}$. Recall that the Dirac equation is invariant under these transformations.}. This representation ensures that we start with an equation of the form of \eqref{eq:gen_form} and \eqref{eq:gen_form2}  with real coefficients. Explicitly, we have for all $x \in [x^{-},x^{+}]$
\begin{eqnarray}
\label{eq:dir_eq_exp_1}
 \frac{E}{c}\psi_{1}(x) &=& \partial_{x} \psi_{2}(x) - mc \psi_{1}(x) + \frac{1}{c}V(x) \psi_{1}(x),\\ 
\label{eq:dir_eq_exp_2}
 \frac{E}{c}\psi_{2}(x) &=& -\partial_{x} \psi_{1}(x) + mc \psi_{2}(x) + \frac{1}{c}V(x) \psi_{2}(x).
\end{eqnarray}
Now, let $u^{+},v^{+}$ and $u^{-},v^{-}$ defined for $ x\in [x_{0},x^{+}]$ and $x \in [x^{-},x_{0}]$ respectively (for $x_{0} \in [x^{-},x^{+}]$), be two solutions of  (\ref{eq:dir_eq_exp_1}) and (\ref{eq:dir_eq_exp_2}) having the following ``initial data'' \cite{Titchmarsh01011961}
\begin{eqnarray}
\label{eq:init_cond_u}
 u_{1}^{\pm}(x_{0}) = 1, \;\; u_{2}^{\pm}(x_{0}) = 0, \\ 
\label{eq:init_cond_v}
 v_{1}^{\pm}(x_{0}) = 0, \;\; v_{2}^{\pm}(x_{0}) = 1 .
\end{eqnarray}
Here, $x_{0}$ can be chosen arbitrarily and from now on, we set $x_{0}=0$. This particular set of ``initial'' data guarantees that the condition $u_{1}^{\pm}(x)v_{2}^{\pm}(x) - u_{2}^{\pm}(x)v_{1}^{\pm}(x) = 1$ is fulfilled for all values of $x$, respectively (it can be shown that the value of this expression does not depend on the space coordinate) and that the solutions $u^{\pm}$ and $v^{\pm}$ are linearly independent. Thus, a set of two eigenfunctions is found with initial conditions (at an arbitrary point within the domain) chosen such that the Wronskian is normalized to 1. Thus, we can write the general solution as a linear combination of these solutions (up to an overall normalization constant) such as 
\begin{eqnarray}
 \psi^{+}(x) = u^{+}(x) + l^{+}(E) v^{+}(x), \;\;x\in [0,x^{+}], \\
 \psi^{-}(x) = u^{-}(x) + l^{-}(E) v^{-}(x), \;\;x \in [x^{-},0],
\end{eqnarray}
where $l^{\pm} \in \mathbb{C}$ are integration constants that need to be fixed by boundary conditions at $x=x^{\pm}$, respectively. The choice of these boundary conditions is very important to ensure the \textit{self-adjointness} of the operator $H$. A general choice is given by 
\begin{eqnarray}
\label{eq:BC_inf}
 \psi_{1}^{\pm}(x^{\pm}) \cos\beta + \psi_{2}^{\pm}(x^{\pm}) \sin\beta = 0  
\end{eqnarray}
where $\beta \in \mathbb{R}$ is an arbitrary parameter. This choice is very similar to the one in Sturm-Liouville problems: it contains the Dirichlet and Neumann conditions as special cases, and it ensures that the operator is self-adjoint. The integration constants obey $l^{\pm}(E) \xrightarrow{x^{\pm} \rightarrow \pm \infty} m^{\pm}(E)$. The value of $m^{\pm}(E)$ such that (\ref{eq:BC_inf}) are fulfilled lies on a circle in the complex plane parametrized by $\beta$ and two outcomes are possible \cite{1962eeaw.bookT,nla.cat-vn2169453}:
\begin{enumerate}
 \item The radius of the circle vanishes (limit-point). 
\item The radius of the circle stays finite (limit-circle).
\end{enumerate}
In the former, it can be proved (WTK theorem) that there exists a unique non-trivial solution (independent of $\beta$) such that $ \psi^{\pm} \in L^{2}(0,\infty) \otimes \mathbb{C}^{2}$ ($L^{2}(-\infty,0)\otimes \mathbb{C}^{2}$, respectively) if $\Im(E) > 0$ \cite{1962eeaw.bookT,nla.cat-vn2169453}. In the following, we consider only this limit-point case since the system under study falls into this category \cite{Titchmarsh01011961}. This can be seen easily by looking at the explicit solutions $\psi^{\pm}$ calculated in the next sections and by noting that $\psi^{+} \not\in L^{2}(0,\infty) \otimes \mathbb{C}^{2}$ and $\psi^{-} \not\in L^{2}(-\infty,0) \otimes \mathbb{C}^{2}$ when $\Im(E)=0$. This, by definition, is the limit-point type \cite{nla.cat-vn2169453}.

The functions $m^{\pm}(E)$ are the Weyl-Titchmarsh $m$-functions and their knowledge allow us to compute the spectral density. The general procedure to compute $m^{\pm}(E)$ is to construct a linear combination of two solutions obeying (\ref{eq:init_cond_u}) and (\ref{eq:init_cond_v}). Then, the functions $m^{\pm}(E)$ are chosen such that the linear combination vanishes at the boundaries when the energy has a non-zero positive imaginary part. For example, if the singularity is at $x=\infty$, one must look at the asymptotic solution and make sure it vanishes in that limit when $\Im(E) >0$, guaranteeing that the solution is in $L^{2}(0,\infty) \otimes \mathbb{C}^{2}$ and obeying the WTK theorem.

The spectral density described earlier in (\ref{eq:eigen_exp}) can be related to the Weyl-Titchmarsh functions by applying the the resolvent operator on $f$ and by comparing with usual results for the Green's function \cite{fulling1989aspects}. The final result is that the spectral density is given by
\begin{eqnarray}
 \boldsymbol{\rho}(E) = \lim_{\epsilon \rightarrow 0^{+}} \frac{1}{\pi} \Im \mathbf{M}(E+i \epsilon)
\end{eqnarray}
where the matrix $\mathbf{M}$ is defined as
\begin{eqnarray}
 \mathbf{M}(E) = \frac{1}{m^{+}(E) - m^{-}(E)}
\begin{bmatrix}
 -1 & \frac{1}{2}\left[ m^{+}(E) + m^{-}(E) \right]\\
\frac{1}{2}\left[ m^{+}(E) + m^{-}(E) \right] & m^{+}(E) m^{-}(E)
\end{bmatrix}.
\end{eqnarray}
However, it is more convenient for our purpose to work with the ``trace spectral density'' given by $\rho = \mathrm{Tr}(\boldsymbol{\rho})$, or more explicitly by \cite{PhysRevA.69.053409,PhysRevLett.81.2663}
\begin{equation}
\label{eq:spec_trace}
 \rho(E) := \lim_{\epsilon \rightarrow 0^{+}} \frac{1}{\pi} \Im \left[ \frac{m^{+}(E+i\epsilon)m^{-}(E+i\epsilon) + 1}{m^{+}(E+i\epsilon) - m^{-}(E+i\epsilon)} \right].
\end{equation}
The latter contains all the physical information on the system, i.e. the whole spectrum is included in this expression \cite{nla.cat-vn2169453}. Note here that if the potential is an even function, the relation $m^{+}(E)=-m^{-}(E)$ holds and we have \cite{Titchmarsh01011961}
\begin{equation}
\label{eq:spec_even}
 \rho(E) = \lim_{\epsilon \rightarrow 0^{+}} \frac{1}{2\pi} \Im \left[ - m^{+}(E+i\epsilon) + \frac{ 1}{m^{+}(E+i\epsilon)} \right].
\end{equation}
Thus, for this specific case, we only have to construct the solution $\psi^{+}$.


\section{Free case}
\label{sec:free}

Now, we consider an explicit example of the theory described in the preceding section: the free case on the real line. This is probably the simplest possible case and the solution can be found in \cite{springerlink:10.1007/BF00401163,Benvegnu1994}, although in the form of the resolvent operator. The reason for presenting this example is twofold: it allows to implement the method in a simple setting and also, it will serve as a validation tool for the results obtained in the next sections. 

\begin{figure}
\centering
\includegraphics[width=0.7\textwidth]{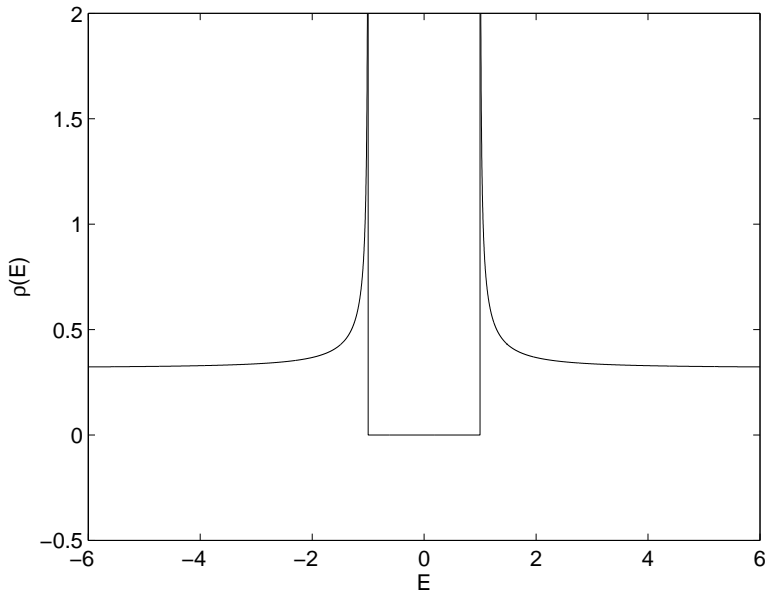}
\caption{Free spectral density.}
\label{fig:rho_free}
\end{figure}

Specifically, the free Dirac equation is given by
\begin{eqnarray}
\label{eq:dir_eq_free_1}
 \frac{E}{c}\psi_{1}(x) &=& \partial_{x} \psi_{2}(x) - mc \psi_{1}(x), \\ 
\label{eq:dir_eq_free_2}
 \frac{E}{c}\psi_{2}(x) &=& -\partial_{x} \psi_{1}(x) + mc \psi_{2}(x). 
\end{eqnarray}
The general solution to this system of equation can be easily computed in terms of trigonometric functions. This solution, for all $x$, is given by
\begin{eqnarray}
 \psi_{1}(x) &=& c_{1} \sin \left( \frac{p}{c} x \right) + c_{2}  \cos \left( \frac{p}{c} x \right) ,\\
 \psi_{2}(x) &=& -c_{1}\frac{p}{E-mc^{2}} \cos \left( \frac{p}{c} x \right) + c_{2}\frac{p}{E-mc^{2}} \sin \left( \frac{p}{c} x \right) ,
\end{eqnarray}
where $p=\sqrt{E^{2} - m^{2}c^{4}}$ (note that we have $p=i \sqrt{m^{2}c^{4} - E^{2}}$ when $|E|<mc^{2}$, that is we chose the Riemann sheet where $\Im (p) \geq 0$) and $c_{1,2}$ are integration constants. These integration constants have to be fixed by using suitable boundary conditions. According to the theory developed in the last section, we are interested in finding solutions obeying the boundary conditions in  (\ref{eq:init_cond_u}) and (\ref{eq:init_cond_v}). We define these solutions as $u$ and $v$, and they are given explicitly for all $x$ by
\begin{eqnarray}
\label{eq:free_sol_1}
u_{1}(x) &=& \cos \left( \frac{p}{c} x \right) ,\\
\label{eq:free_sol_2}
u_{2}(x) &=& \frac{p}{E-mc^{2}} \sin \left( \frac{p}{c} x \right) ,\\
\label{eq:free_sol_3}
v_{1}(x) &=& -\frac{E-mc^{2}}{p} \sin \left( \frac{p}{c} x \right), \\
\label{eq:free_sol_4}
v_{2}(x) &=& \cos \left( \frac{p}{c} x \right).
\end{eqnarray}
From these solutions, we now form the general solution $\psi^{+}$ (in the free case, the potential is even, so $\psi^{-}$ is not needed). It is given by
\begin{eqnarray}
\label{eq:psi_plus_free}
 \psi^{+}(x) = u(x) + m^{+}(E)v(x), \;\; x \in \mathbb{R}^{+}.
\end{eqnarray}
For general $m^{+}$ however, this solution is not in $L^{2}(0,\infty) \otimes \mathbb{C}^{2}$ when $\Im(E)>0$, as can be seen by looking at the function behavior at infinity: asymptotically, we have $  \psi^{+}(x) \underset{\lim_{x \rightarrow + \infty}}{\sim} e^{\pm i \frac{p}{c}x}$. However, by following the WTK prescription and giving a small positive imaginary part $\epsilon$ to the eigenenergy, we have
\begin{eqnarray}
p=\sqrt{(E+i \epsilon)^{2} - m^{2}c^{4}} = \sqrt{E^{2} - m^{2}c^{4}} + i\frac{E}{\sqrt{E^{2} - m^{2}c^{4}}} \epsilon + O(\epsilon^{2})
\end{eqnarray}
and thus, there are three possible cases:
\begin{enumerate}
 \item If $E \geq mc^{2}$, then 
\begin{eqnarray}
 \lim_{x \rightarrow +\infty} e^{i\frac{p}{c}x} = 0 \;\; \mbox{and} \;\; \lim_{x \rightarrow +\infty} e^{-i\frac{p}{c}x} = \infty 
\end{eqnarray} 
 \item If $E \leq -mc^{2}$, then 
\begin{eqnarray}
 \lim_{x \rightarrow +\infty} e^{i\frac{p}{c}x} = \infty \;\; \mbox{and} \;\; \lim_{x \rightarrow +\infty} e^{-i\frac{p}{c}x} = 0 
\end{eqnarray} 
 \item If $-mc^{2} < E < mc^{2}$, then 
\begin{eqnarray}
 \lim_{x \rightarrow +\infty} e^{i\frac{p}{c}x} = 0 \;\; \mbox{and} \;\; \lim_{x \rightarrow +\infty} e^{-i\frac{p}{c}x} = \infty 
\end{eqnarray} 
\end{enumerate}
Ensuring that the diverging exponentials are canceled can be done by choosing the Weyl-Titchmarsh $m$-function to be
\begin{eqnarray}
m^{+}(E)= \left\{
\begin{matrix}
 -i \frac{p}{E-mc^{2}}  ,& \mbox{for}& E \geq mc^{2}\\
 i\frac{p}{E-mc^{2}}  ,& \mbox{for}& E \leq -mc^{2}
\end{matrix}
\right. .
\end{eqnarray}
The spectral density of the free case can then be easily computed from  (\ref{eq:spec_even}) and we get
\begin{eqnarray}
 \rho(E) = \left\{
\begin{matrix}
 \frac{1}{\pi} \frac{|E|}{p} ,& \mbox{for}& |E| \geq mc^{2} \\
 0 ,& \mbox{for}&  |E| < mc^{2} \\
\end{matrix}
\right. .
\end{eqnarray}
This spectral density is plotted in Fig. \ref{fig:rho_free}. Note the appearance of the negative energy continuum for $E \leq -mc^{2}=-1$. 

\section{Double delta potential without electric field}
\label{sec:free_delta}

The next case considered is the one of a double-delta potential without electric field. This represents physically a simple model for the ground state (symmetric) and first excited state (anti-symmetric) of a diatomic molecule. We recall to the reader that the same system was treated non-relativistically in \cite{PhysRevA.69.053409}. In this case, the spectrum has a continuum component for $E>0$ while bound states appear as Dirac delta peaks at the bound state energies. Similar features are found in the relativistic case. The main differences however are the appearance of the negative energy continuum while the bound states are positioned in the mass gap $(-mc^{2},mc^{2})$.  

Specifically, the main goal is to compute the spectral density for a potential given by
\begin{eqnarray}
\label{eq:pot_2dirac_free}
 V(x) = - g \left[ \delta(x-R) + \delta(x+R) \right] , \;\; x \in \mathbb{R}
\end{eqnarray}
where $2R$ is the interatomic distance and $g$ is the strength of the Dirac delta potential wells. 

The calculation proceeds as in the free case: the first step is to find two solutions obeying the boundary conditions (\ref{eq:init_cond_u}) and (\ref{eq:init_cond_v}), then one constructs a general solution in $L^{2}(0,\infty) \otimes \mathbb{C}^{2}$ ($\psi^{+}$) from which we can obtain the $m$-functions and thus, the spectral density. 

The main difference with the free case is the treatment of the point interaction. As usual with these kinds of potential, the solution can be found by solving the free Dirac equation by including boundary conditions at $x = \pm R$. However, it is well-known that there is an ambiguity in the definition of the boundary condition corresponding to the potential in  (\ref{eq:pot_2dirac_free}) in the Dirac equation \cite{PhysRevA.24.1194}. This ambiguity can be traced to the fact that the wave function should have a jump continuity at $x = \pm R$ and thus, the terms $\psi(x) \delta(x \pm R)$ appearing in the equation are actually products of distributions (or generalized function). The appearance of the jump discontinuity in the wave function can be understood heuristically in the following way, as suggested in \cite{calkin:737}. Let us consider a massless Dirac equation at $E=0$ (a finite $E$ does not change the argument) and only one Dirac delta potential positioned at $x=0$ such that $-i\alpha \partial_{x} \psi(x) = \delta(x)\psi(x)$. For this equality to hold, the wave function has to be of the form $\psi \sim H$ (where $H$ is the Heaviside function) which implies a jump discontinuity. 

The strategy used in this work to obtain the boundary conditions is to formulate everything in the weak sense and consider the product of the potential and wave function as a product of distributions from the outset. The main problem is that the set of distributions is not an algebra \cite{schwartz1966theorie}. There exists however a mathematical construction that includes these product of distributions rigorously: the Colombeau's theory of generalized function \cite{Colombeau198396,Colombeau1990,colombeau:315,Colombeau:1985}. In that framework, the value of a distribution product depends on the ``local'' structure of the distributions considered, i.e. the choice of the limiting function. Using these ideas, we derive in Appendix \ref{sec:app_bound} a family of boundary conditions consistent with  (\ref{eq:gen_bound}) and thus, having the right mathematical properties (self-adjointness and non-relativistic limit).  
%
%
From this analysis, the following boundary conditions are obtained:
\begin{eqnarray}
\label{eq:bound_cond1_t}
 \psi_{1}(\pm R^{+}) &=& \left[ 1+\frac{g^{2}}{4c^{2}} \right]^{-1} \left[ \left( 1- \frac{g^{2}}{4c^{2}} \right) \psi_{1}(\pm R^{-}) - \frac{g}{c} \psi_{2}(\pm R^{-})  \right] ,\\
\label{eq:bound_cond2_t}
\psi_{2}(\pm R^{+}) &=& \left[ 1+\frac{g^{2}}{4c^{2}} \right]^{-1} \left[ \left( 1- \frac{g^{2}}{4c^{2}} \right) \psi_{2}(\pm R^{-}) + \frac{g}{c} \psi_{1}(\pm R^{-})  \right],
\end{eqnarray}
and 
\begin{eqnarray}
\label{eq:bound_cond3_t}
 \psi_{1}(\pm R^{-}) &=& \left[ 1+\frac{g^{2}}{4c^{2}} \right]^{-1} \left[ \left( 1- \frac{g^{2}}{4c^{2}} \right) \psi_{1}(\pm R^{+}) + \frac{g}{c} \psi_{2}(\pm R^{+})  \right], \\
\label{eq:bound_cond4_t}
\psi_{2}(\pm R^{-}) &=& \left[ 1+\frac{g^{2}}{4c^{2}} \right]^{-1} \left[ \left( 1- \frac{g^{2}}{4c^{2}} \right) \psi_{2}(\pm R^{+}) - \frac{g}{c} \psi_{1}(\pm R^{+})  \right],
\end{eqnarray}
where we defined $\psi(\pm R^{+}) = \lim_{\epsilon \rightarrow 0} \psi(\pm R + \epsilon)$ and $\psi(\pm R^{-}) = \lim_{\epsilon \rightarrow 0} \psi(\pm R - \epsilon)$.

\begin{figure}
\centering
 \includegraphics[width=0.7\textwidth]{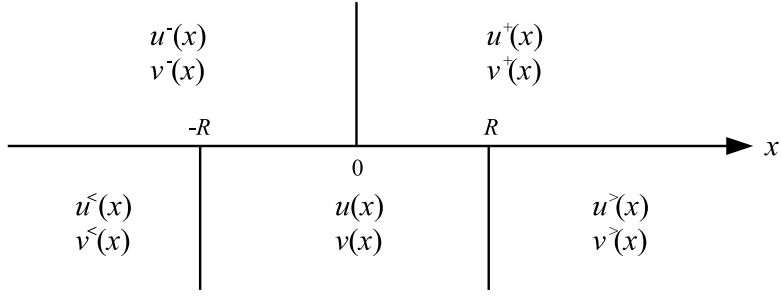}
 \caption{Domain of the solutions.}
 \label{fig:dom_sol}
 \end{figure}

Having now all the necessary elements, it is possible to compute the spectral density for the double Dirac system without electric field. The procedure is now summarized:
\begin{enumerate}
 \item Compute $u$ and $v$ for $ x \in [0,R)$, which are solutions obeying the boundary conditions  (\ref{eq:gen_bound}) and (\ref{eq:gen_bound2})).
\item Compute $u^{>}$ and $v^{>}$ for $ x \in (R,\infty)$ using the conditions  (\ref{eq:bound_cond1_t}) and (\ref{eq:bound_cond2_t}). This yields the solutions on $x \in [0,\infty)$ as (see Figure \ref{fig:dom_sol} for the domain of each function)
\begin{eqnarray}
 u^{+}(x) &=& \left\{
\begin{matrix}
 u(x) & \mbox{for} & 0<x<R \\
u^{>}(x) & \mbox{for} & R<x<\infty
\end{matrix} \right. ,\\
 v^{+}(x) &=& \left\{
\begin{matrix}
 v(x) & \mbox{for} & 0<x<R \\
v^{>}(x) & \mbox{for} & R<x<\infty
\end{matrix} \right. .
\end{eqnarray}
\item Compute $m^{+}$ by ensuring that $\psi^{+}(x) = u^{+}(x) + m^{+}v^{+}(x)$ is in $L^{2}(0,\infty) \otimes \mathbb{C}^{2}$ when $\Im (E) >0$. 
\item Compute $\rho(E)$ from $m^{+}$.
\end{enumerate}

\subsection{Solution in terms of trigonometric functions}

The first part of the calculation was already performed in the free case. In the latter, the free Dirac equation,  (\ref{eq:dir_eq_free_1}) and (\ref{eq:dir_eq_free_2}), was solved with suitable boundary conditions. The solution obtained in  (\ref{eq:free_sol_1}) to (\ref{eq:free_sol_4}) thus corresponds to the needed solution in the region $x \in [0,R)$, i.e. $u(x)$ and $v(x)$.  

The next step is to compute the wave functions in the region $x>R$. They are given by
\begin{eqnarray}
 u_{1}^{>}(x) &=& a_{1} \sin \left( \frac{p}{c} x \right) + a_{2}  \cos \left( \frac{p}{c} x \right) ,\\
 u_{2}^{>}(x) &=& -a_{1}\frac{p}{E-mc^{2}} \cos \left( \frac{p}{c} x \right) + a_{2}\frac{p}{E-mc^{2}} \sin \left( \frac{p}{c} x \right) , \\
 v_{1}^{>}(x) &=& b_{1} \sin \left( \frac{p}{c} x \right) + b_{2}  \cos \left( \frac{p}{c} x \right) ,\\
 v_{2}^{>}(x) &=& -b_{1}\frac{p}{E-mc^{2}} \cos \left( \frac{p}{c} x \right) + b_{2}\frac{p}{E-mc^{2}} \sin \left( \frac{p}{c} x \right) .
\end{eqnarray}
The value of the integration constants $a_{1},a_{2},b_{1}$ and $b_{2}$ can be determined from the boundary conditions in (\ref{eq:bound_cond1_t}) and (\ref{eq:bound_cond2_t}). These conditions yield the following relations:
\begin{eqnarray}
u_{1}^{>}(R) &=& \left[ 1+\frac{g^{2}}{4c^{2}} \right]^{-1} \left[ \left(1-\frac{g^{2}}{4c^{2}}  \right) u_{1}(R) - \frac{g}{c} u_{2}(R)  \right] := G_{u}^{-} ,\\
u_{2}^{>}(R) &=& \left[ 1+\frac{g^{2}}{4c^{2}} \right]^{-1} \left[ \left(1-\frac{g^{2}}{4c^{2}}  \right) u_{2}(R) + \frac{g}{c} u_{1}(R)  \right] := G_{u}^{+} ,\\
v_{1}^{>}(R) &=& \left[ 1+\frac{g^{2}}{4c^{2}} \right]^{-1} \left[ \left(1-\frac{g^{2}}{4c^{2}}  \right) v_{1}(R) - \frac{g}{c} v_{2}(R)  \right] := G_{v}^{-},\\
v_{2}^{>}(R) &=& \left[ 1+\frac{g^{2}}{4c^{2}} \right]^{-1} \left[ \left(1-\frac{g^{2}}{4c^{2}}  \right) v_{2}(R) + \frac{g}{c} v_{1}(R)  \right] := G_{v}^{+}.
\end{eqnarray}
As usual, it is possible to solve for the coefficients and we get
\begin{eqnarray}
a_{1} &=&   G_{u}^{-} \sin\left( \frac{p}{c} R \right) - \frac{E-mc^{2}}{p}G_{u}^{+} \cos\left( \frac{p}{c} R \right), \\
a_{2} &=& \frac{E-mc^{2}}{p}  G_{u}^{+} \sin\left( \frac{p}{c} R \right) + G_{u}^{-} \cos\left( \frac{p}{c} R \right), \\
b_{1} &=& G_{v}^{-} \sin\left( \frac{p}{c} R \right) - \frac{E-mc^{2}}{p}G_{v}^{+} \cos\left( \frac{p}{c} R \right) ,\\
b_{2} &=& \frac{E-mc^{2}}{p}  G_{v}^{+} \sin\left( \frac{p}{c} R \right) + G_{v}^{-} \cos\left( \frac{p}{c} R \right) .
\end{eqnarray}
This completes the derivation of the solution on $x \in [0,\infty)$ for $u^{+}$ and $v^{+}$. The solution on $x<0$ could be calculated in a similar way but can be easily obtained from $\psi^{+}$ owing to the symmetry of the potential. The next step is the evaluation of the $m$-function.

\subsection{Evaluation of the Weyl-Titchmarsh $m$-function}

In this section, the function $m^{+}$ is evaluated. Remember that the general solution $\psi^{+} = u^{+} + m^{+}(E)v^{+}$ should be in $L^{2}(0,\infty) \otimes \mathbb{C}^{2}$ when $\Im (E) >0$. A similar calculation was carried in the free case so the details are not repeated here. The final result is that 
\begin{eqnarray}
 m^{+}(E)= \left\{ 
\begin{matrix}
 - \frac{-ia_{1} + a_{2} }{-ib_{1} +b_{2}} & \mbox{for}& E< -mc^{2}\\
 - \frac{ia_{1} + a_{2} }{ib_{1} + b_{2}} & \mbox{for}& E> -mc^{2}
\end{matrix}
 \right. .
\end{eqnarray}
This choice ensures that the diverging exponentials are canceled as $x \rightarrow \infty$ and $\Im(E) > 0$. Thus, by combining this last equation with the solution computed in the last section, we can evaluate the spectral density.

\subsection{Evaluation of the spectral density}

\begin{figure}
\subfloat[]{\includegraphics[width=3.0in]{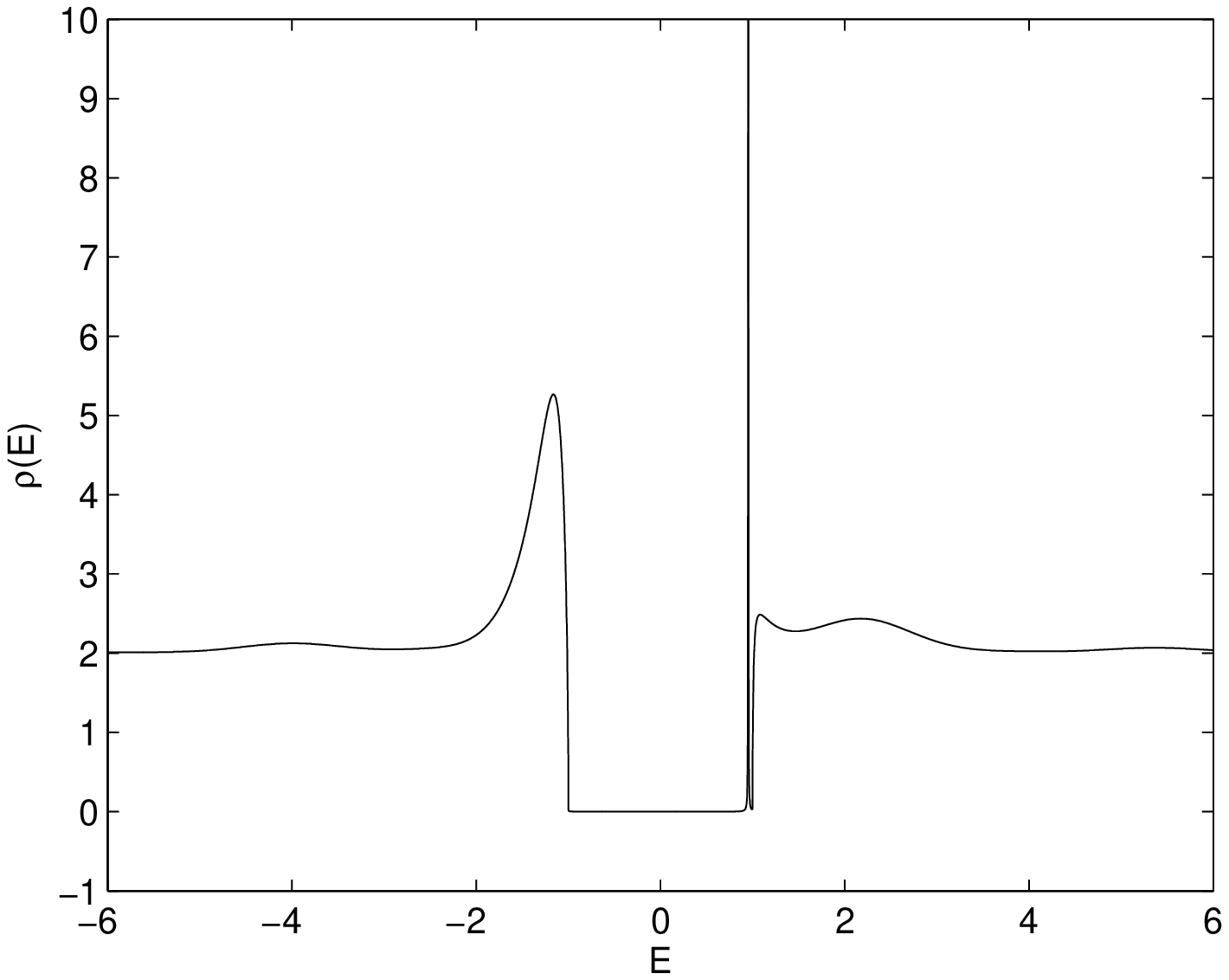}}
\subfloat[]{\includegraphics[width=3.0in]{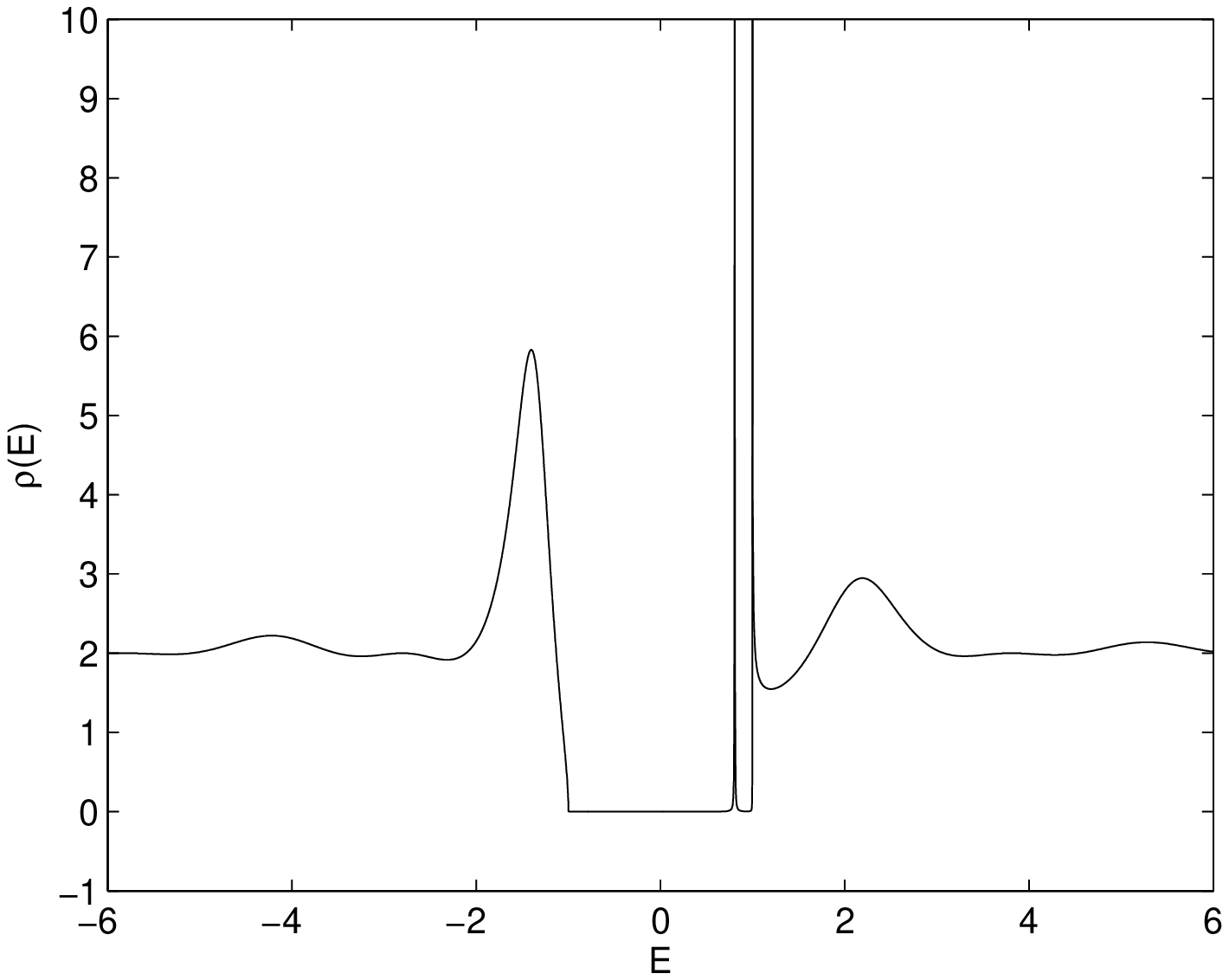}} \\
\subfloat[]{\includegraphics[width=3.0in]{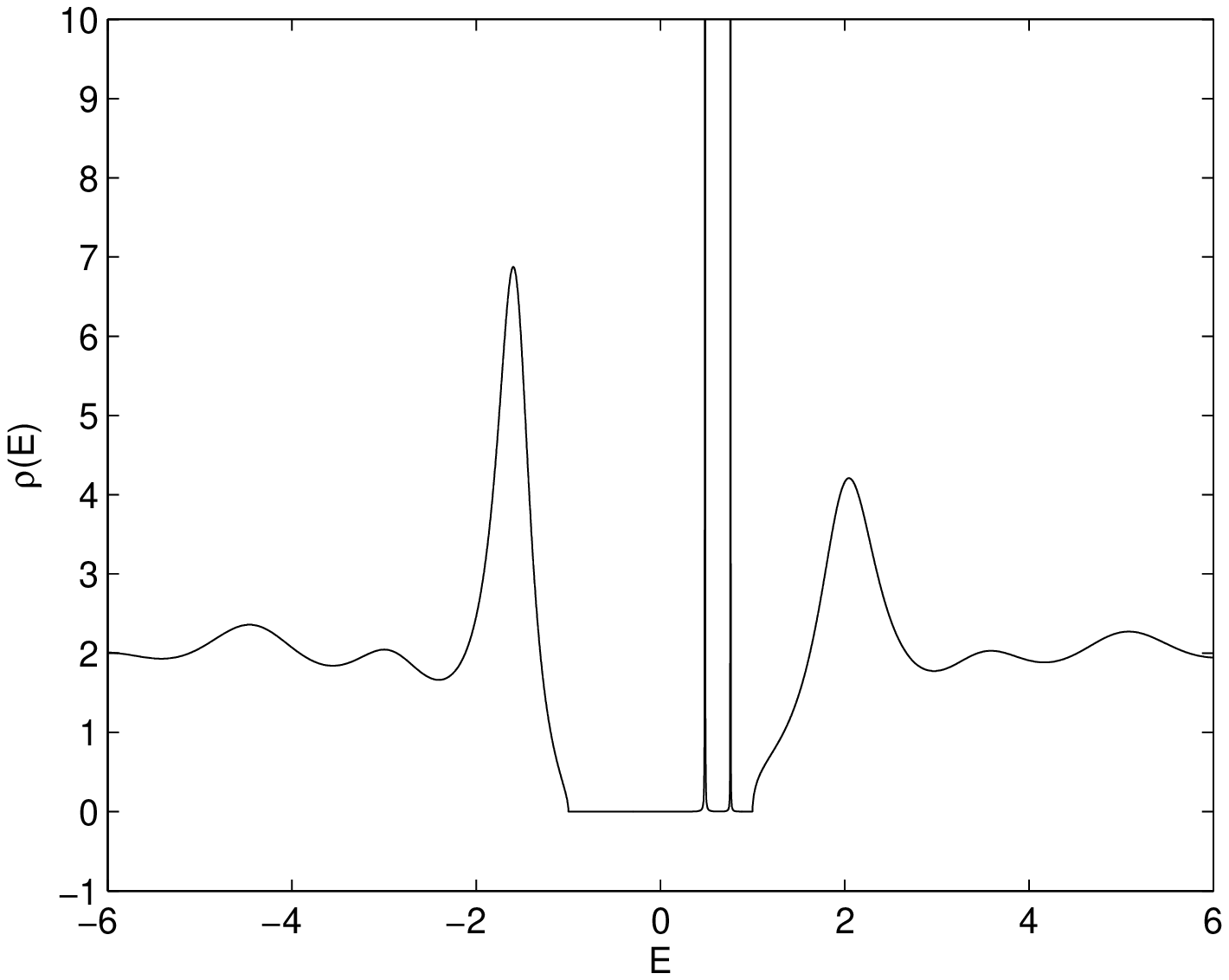}}
\subfloat[]{\includegraphics[width=3.0in]{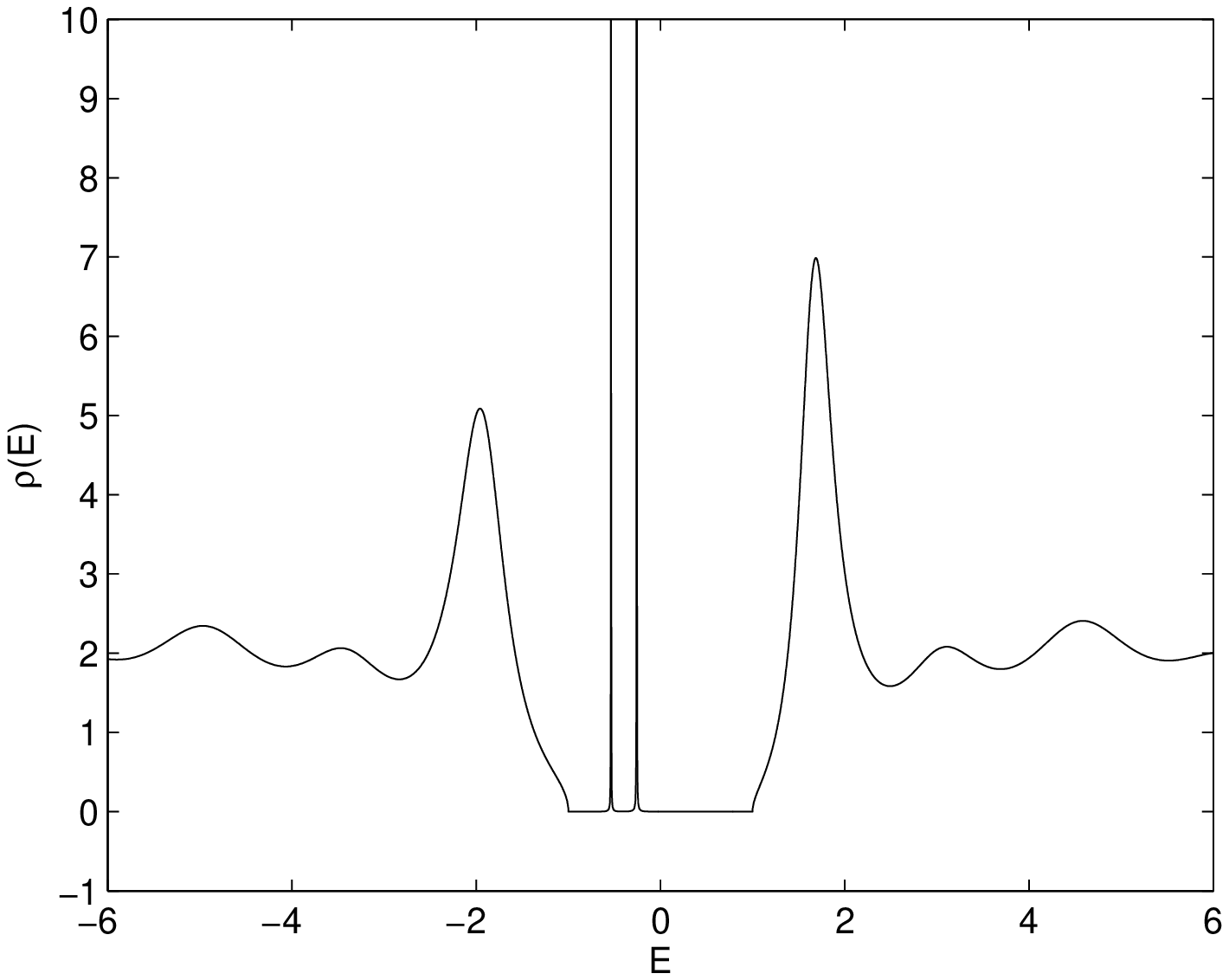}} \\
\subfloat[]{\includegraphics[width=3.0in]{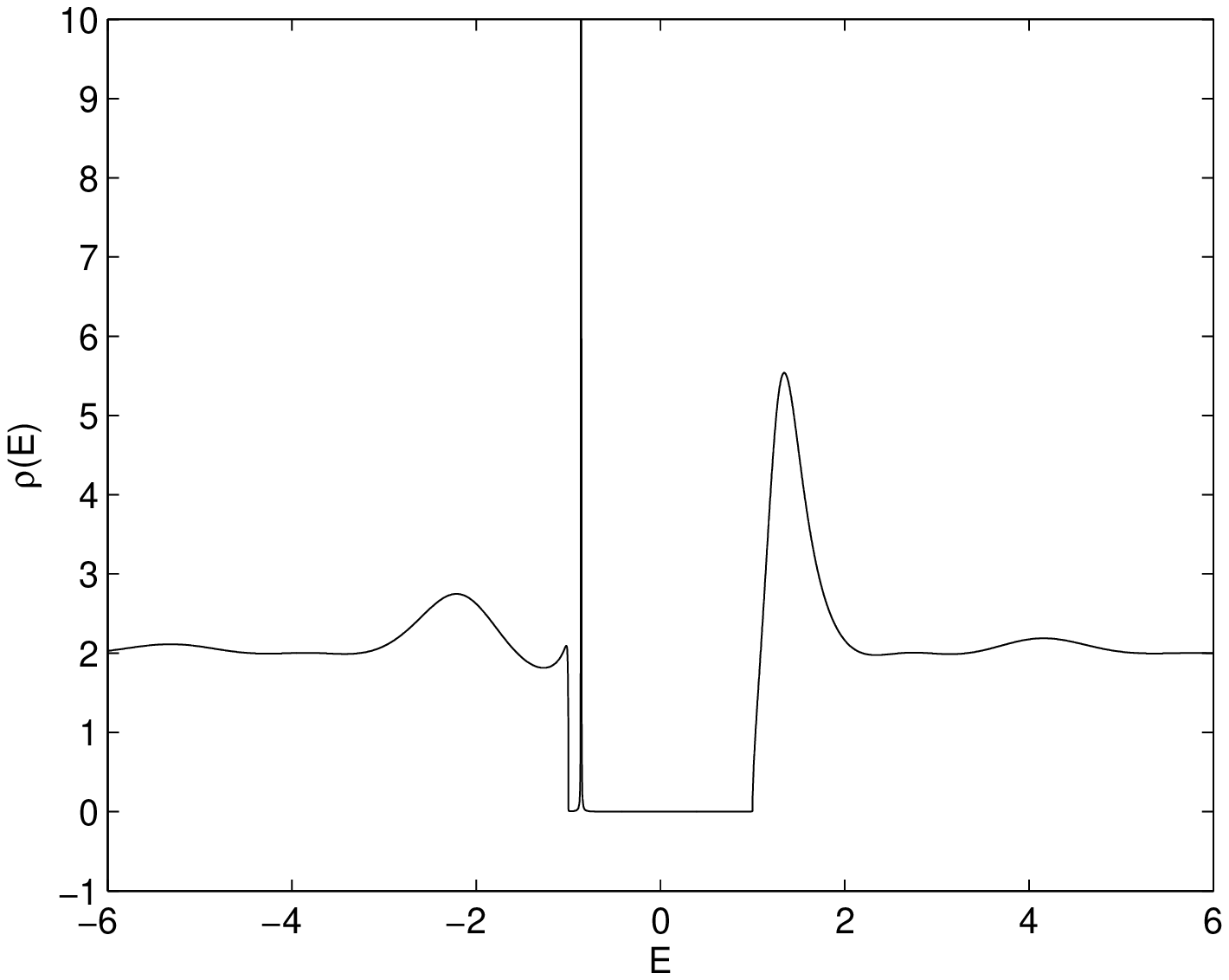}}
\subfloat[]{\includegraphics[width=3.0in]{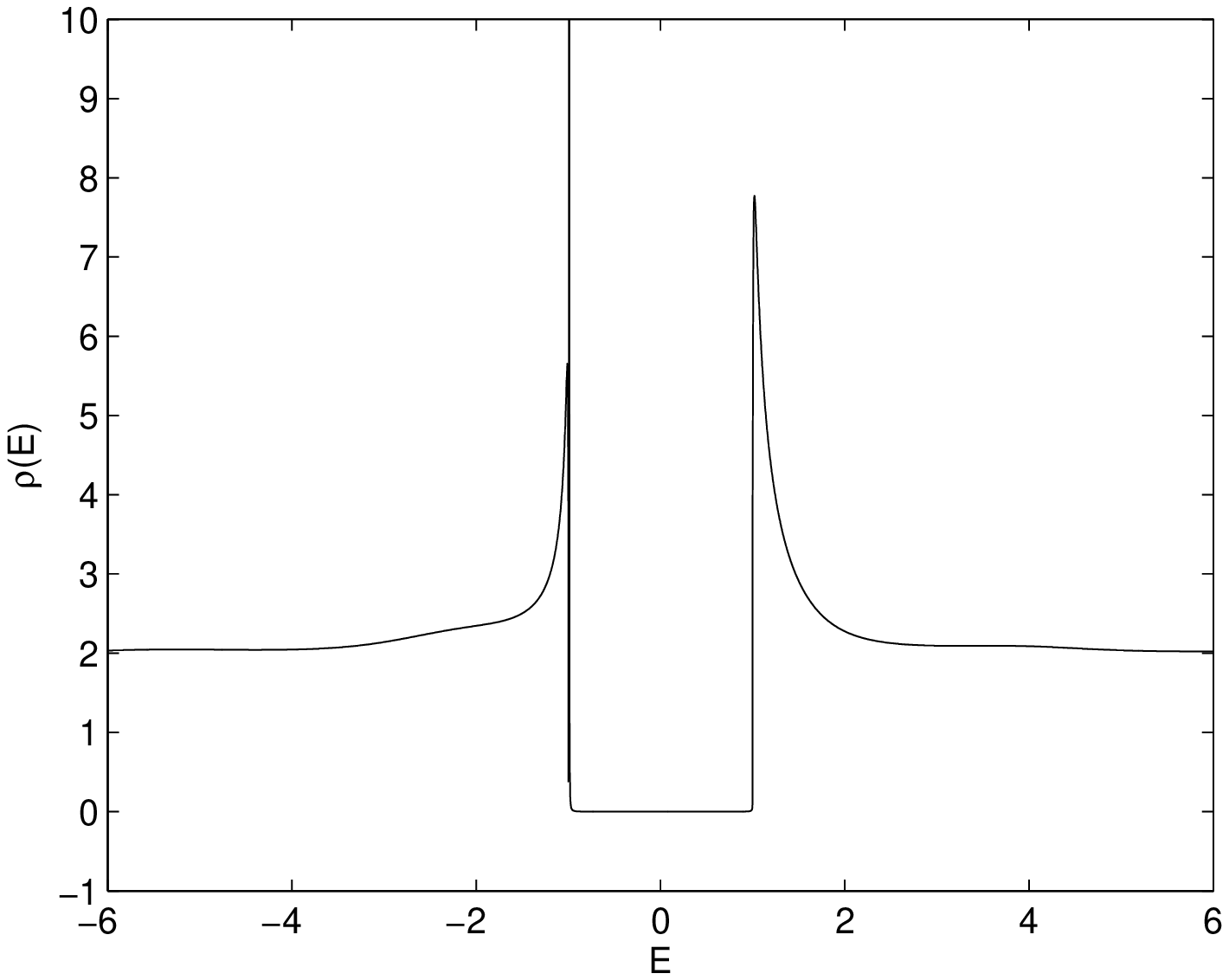}}  
\caption{Spectral density for a two Dirac delta potential for varying values of $g$. The value of the interatomic distance is fixed to $R=1.0$ while the potential strength is (a) $g=0.2$, (b) $g=0.5$, (c) $g=1.0$, (d) $g=3.0$, (e) $g=10.0$ and (f) $g=50.0$.}
\label{fig:rho_2_del_g}
\end{figure}

\begin{figure}
\subfloat[]{\includegraphics[width=3.0in]{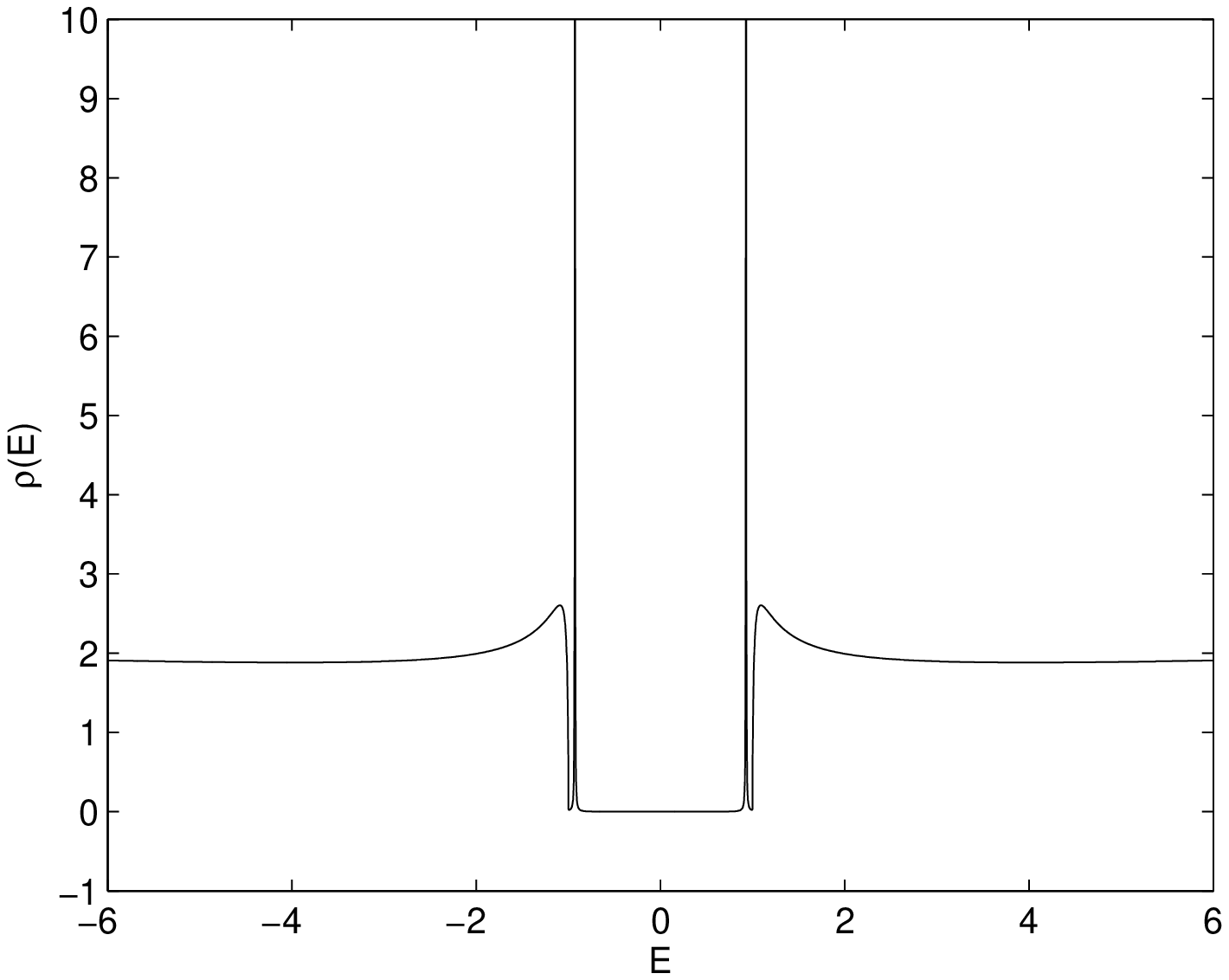}}
\subfloat[]{\includegraphics[width=3.0in]{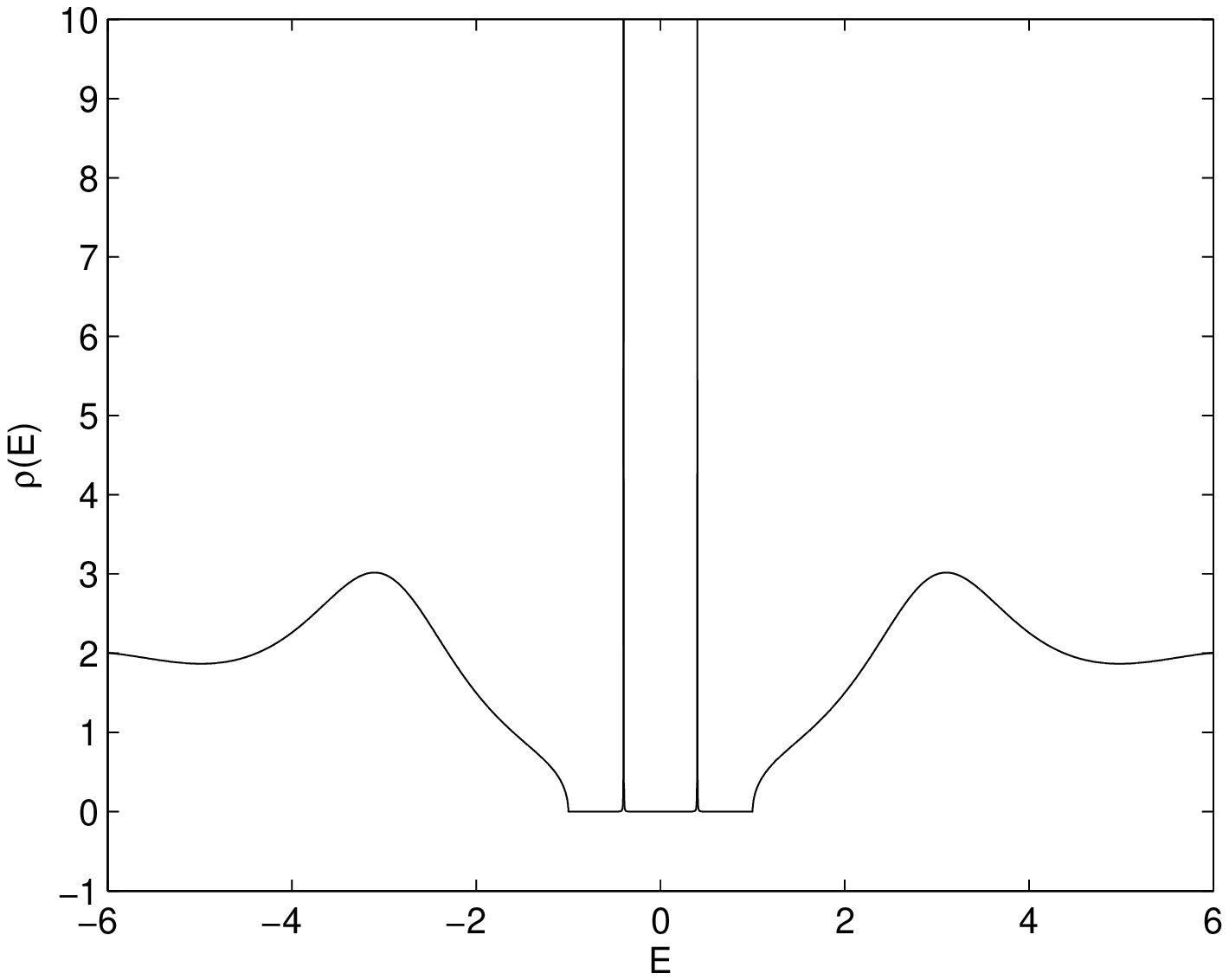}} \\
\subfloat[]{\includegraphics[width=3.0in]{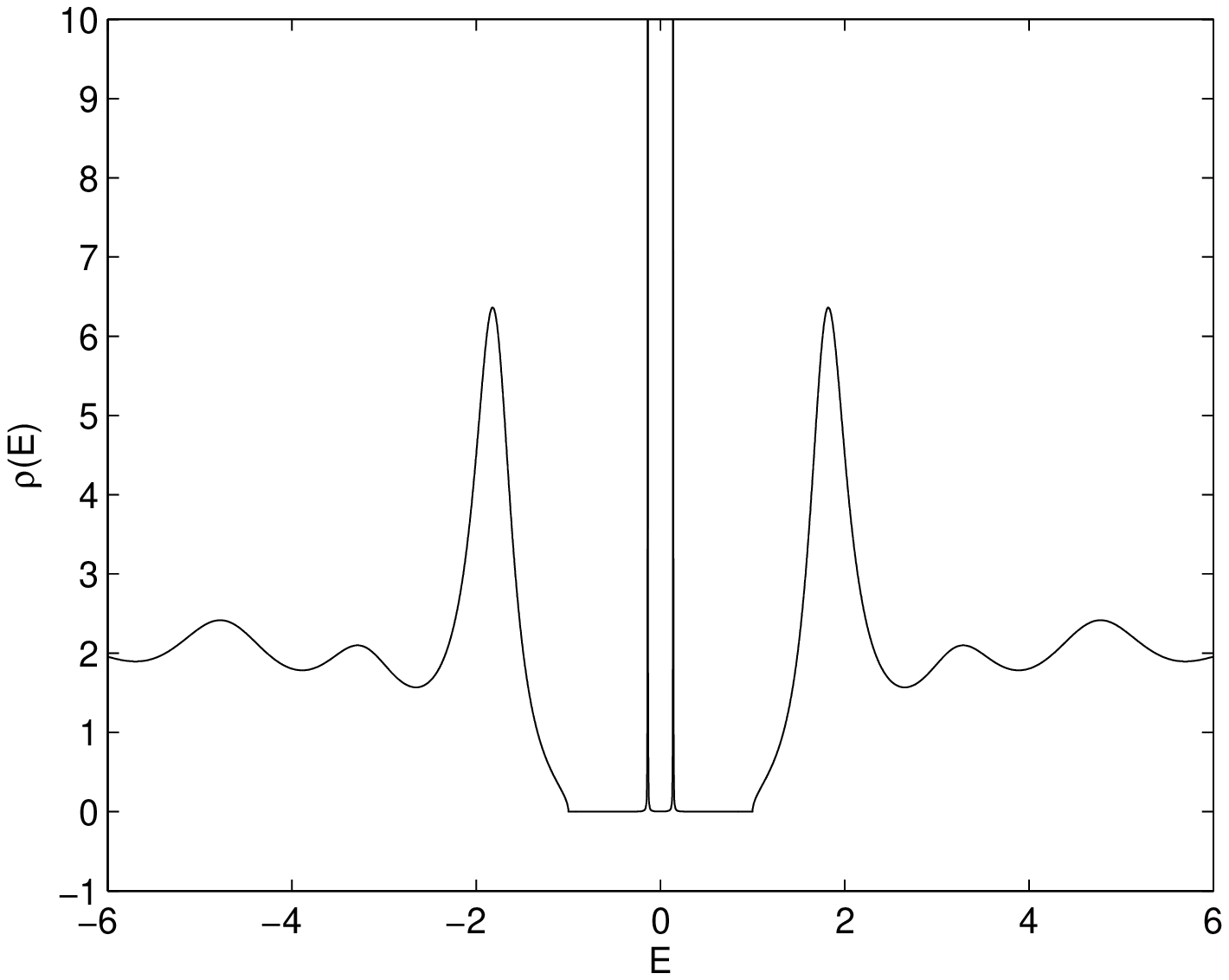}} 
\subfloat[]{\includegraphics[width=3.0in]{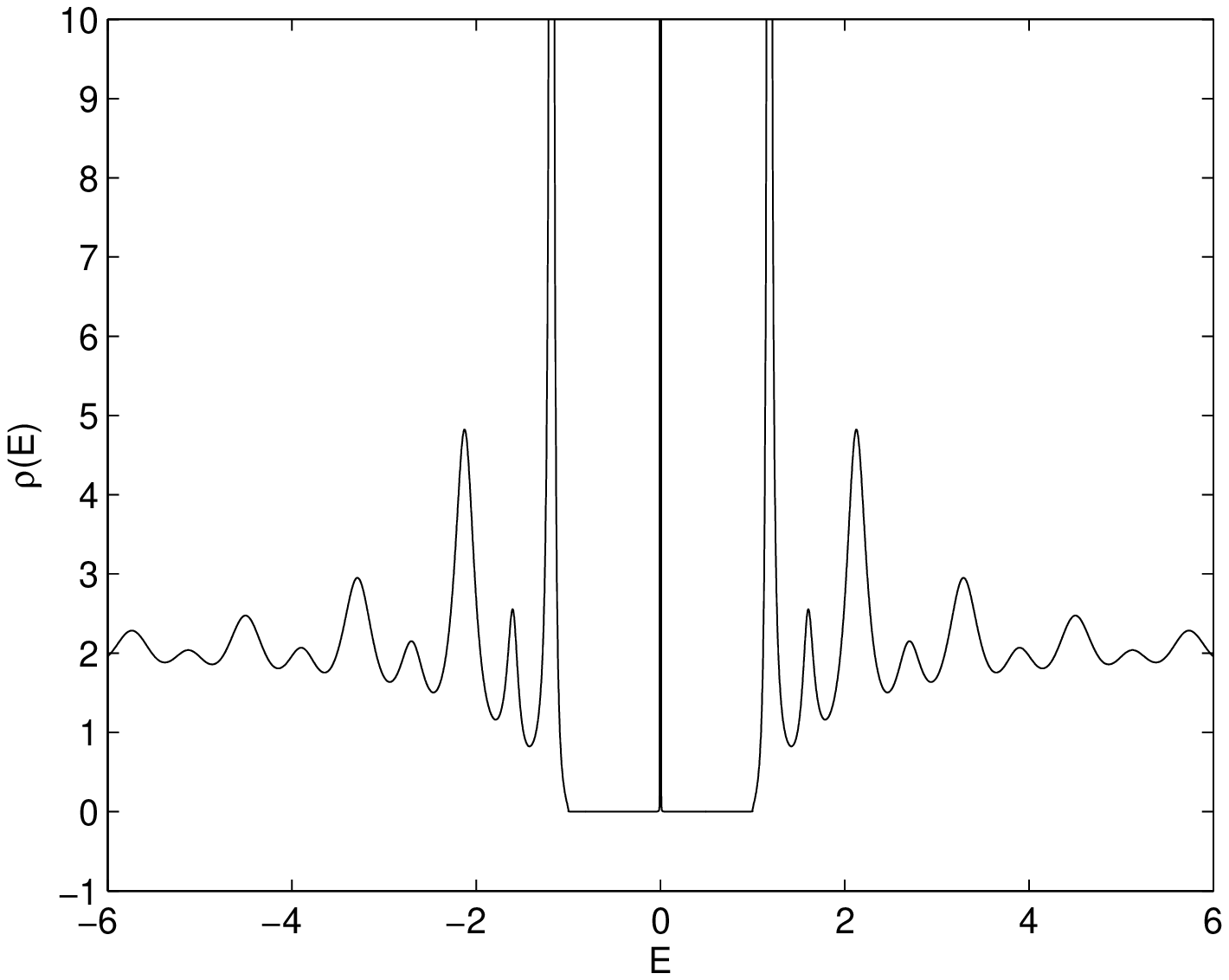}}
\caption{Spectral density for a two Dirac delta potential for varying values of $R$. The value of the potential strength is fixed to $g=2.0$ while the interatomic distance is (a) $R=0.1$, (b) $R=0.5$, (c) $R=1.0$ and (d) $R=2.5$.}
\label{fig:rho_2_del_R}
\end{figure}

The spectral density is given by  (\ref{eq:spec_even}) when the potential is even. Having determined the $m$-function in the last section, it is now possible to evaluate the spectral density. It is a complicated function of the interatomic distance and potential strength, so a numerical evaluation seems better suited than showing the equation. However, the bound state energies have relatively simple expressions: they are given by the position of the poles in the spectral density and thus, satisfy the following transcendental equation
\begin{eqnarray}
\label{eq:pole_pos}
 -\left(1 - \frac{g^{2}}{4c^{2}} \right) \sqrt{m^{2}c^{4} - E^{2}} + \frac{g}{c} \left[ E \pm mc^{2} e^{-2\tilde{p}R} \right] = 0
\end{eqnarray}
for the energy $E$ having a value in the mass gap, that is $E \in [-mc^{2} , mc^{2}]$. Here, we defined $\tilde{p} = \sqrt{m^{2}c^{4} - E^{2}}$ and the cases $\pm$ correspond to the ground and first excited states respectively. It was verified that the solutions of  (\ref{eq:pole_pos}) corresponds to the position of the bound state in Figs. \ref{fig:rho_2_del_g} and \ref{fig:rho_2_del_R}.  

In Fig. \ref{fig:rho_2_del_g}, the spectral density is plotted for different values of the potential strength ($g$) and a fixed atomic interatomic distance ($R=1$). As shown in the figure, there is only one bound state at low potential strength (the ground state). When the potential strength reaches $g = -4mc^{3}R + 2c \sqrt{4m^{2}c^{4}R^{2}+1}$, the second bound state appears (the first excited state), while the ground state has a lower energy. By increasing $g$ furthermore, the two bound states have lower eigenenergies until the ground state disappears in the negative energy continuum when $g = 4mc^{3}R + 2c \sqrt{4m^{2}c^{4}R^{2}+1}$. To summarize, we have
\begin{enumerate}
 \item For $0 \leq g <-4mc^{3}R + 2c \sqrt{4m^{2}c^{4}R^{2}+1}$, the ground state is the only bound state in the mass gap.
\item For $-4mc^{3}R + 2c \sqrt{4m^{2}c^{4}R^{2}+1} \leq g<4mc^{3}R + 2c \sqrt{4m^{2}c^{4}R^{2}+1}$, both the ground and the excited states are bound states in the mass gap.
\item For $4mc^{3}R + 2c \sqrt{4m^{2}c^{4}R^{2}+1} \leq g$, the excited state is the only bound state in the mass gap, the ground state has vanished in the negative energy continuum.
\end{enumerate}
Note here that for $g \sim 0$ or $g \gg 1$, the free spectral density is recovered.  

In Fig. \ref{fig:rho_2_del_R}, the spectral density is shown for many values of interatomic distance ($R$) and for fixed value of potential strength ($g=2.0$). For small value of $R$, the energy difference between the ground and excited states is larger. As the interatomic distance increases, the ground state eigenenergy also increases while the excited state eigenenergy diminishes until they merge and form a quasi-degenerate state (at large $R$). Note also the qualitative change in the positive and negative continua: as $R$ gets larger, the oscillating behavior is amplified. This phenomenon was interpreted physically in \cite{PhysRevA.69.053409} as the resonant backscattering between the two potential wells (the so-called Ramsauer-Townsend resonances). 

\section{Double delta potential with electric field}

In this section, we consider a more interesting system: the double delta function with an electric field. This system can be seen as a very simplified model of a diatomic molecule in a slowly varying electromagnetic field. Specifically, the main goal is to compute the spectral density for a potential given by
\begin{eqnarray}
 V(x) = - g \left[ \delta(x-R) + \delta(x+R) \right] - F x, \;\; x \in \mathbb{R}
\end{eqnarray}
where $2R$ is the interatomic distance, $F$ is the electric field and $g$ is the strength of the Dirac delta potential wells. Clearly, this potential is singular at $x=\pm \infty$ and therefore, the general theory of eigenfunction expansion should be used. 

As described above, the first step is to find two solutions obeying the boundary conditions  (\ref{eq:init_cond_u}) and (\ref{eq:init_cond_v}). To do this, we start by finding the general solution in region $-R<x<R$ and $x<-R$ and $x>R$, i.e. when there is no delta wells. Once we know the general solution, the case with delta wells can be easily treated by using suitable boundary conditions at $x=\pm R$. The boundary conditions on the point interaction are described in Appendix \ref{sec:app_bound}.

\subsection{Solution in terms of Parabolic Cylinder functions}

The general solution in a constant electric field is known, it was solved in \cite{Titchmarsh01011961}, but we recapitulate the main steps here for completion. A similar calculation was also performed in \cite{GonzalezDiaz2006202}. Following \cite{Titchmarsh01011961}, we start by rewriting the spinor components as 
\begin{eqnarray}
 \Omega_{1} = \frac{1}{\sqrt{2}} \left( \psi_{1} + i \psi_2 \right), \;\;  \Omega_{2} = \frac{1}{\sqrt{2}} \left( \psi_{1} - i \psi_2 \right),
\end{eqnarray}
such that
\begin{eqnarray}
 \psi_{1} = \frac{1}{\sqrt{2}} \left( \Omega_{1} +  \Omega_2 \right), \;\;  \psi_{2} = -\frac{i}{\sqrt{2}} \left( \Omega_{1} -  \Omega_2 \right),
\end{eqnarray}
Substituting the latter in  (\ref{eq:dir_eq_exp_1}) and (\ref{eq:dir_eq_exp_2}), adding and subtracting the two equations, we find that
\begin{eqnarray}
\label{eq:omega1}
 -\left[ \frac{E-V(x)}{c} \right] \Omega_1(x) - i \partial_{x}\Omega_1(x) - mc \Omega_2(x) = 0 ,\\
\label{eq:omega2}
- \left[ \frac{E-V(x)}{c} \right] \Omega_2(x) + i \partial_{x}\Omega_2(x) - mc \Omega_1(x) = 0.
\end{eqnarray}
Then, solving for $\Omega_2$ in  (\ref{eq:omega1}) and substituting in  (\ref{eq:omega2}), we get, after a few manipulations
\begin{eqnarray}
\Omega_{2}(x) = -\frac{1}{mc} \left[ \frac{E+Fx}{c} + i \partial_{x} \right] \Omega_{1}(x) ,\\
 \partial_{x}^{2} \Omega_{1}(x) + \left[ \left( \frac{E+Fx}{c}\right)^{2} - i\frac{F}{c} - m^2 c^2  \right] \Omega_{1}(x) = 0 .
\end{eqnarray}
The second equation can be simplified further by letting $ y = e^{-i\frac{\pi}{4}}\sqrt{\frac{2c}{F}} \left( \frac{E+Fx}{c} \right)$, and this yields
\begin{eqnarray}
 \partial_{y}^{2} \Omega_{1}(y) - \left[ \frac{1}{4}y^{2}  + a  \right] \Omega_{1}(y) = 0 ,
\end{eqnarray}
where $a = i\frac{m^2 c^3}{2F} - \frac{1}{2}$. The last equation has well-known solutions and can involve any pairs of independent solutions among $U(a,\pm z)$,$V(a,\pm z)$ or $U(-a,\pm iz)$ where $U$ and $V$ are the parabolic cylinder functions \cite{AS65,DLMF}. However, it is convenient to use a ``numerically satisfactory pair of solution'', that is a pair for which one solution is dominant while the other is recessive as $\Re (z) \rightarrow \pm \infty$. This facilitates the calculation of the $m$-functions. In our case, the phase of the argument is $\arg \; y = -\pi/4$ and thus, a numerically satisfactory pair of solution is given by \cite{DLMF}
\begin{eqnarray}
 \Omega_{1}(y) = c_{1} U(a,y) + c_{2} U(-a,-iy),
\end{eqnarray}
where $U$ is the parabolic cylinder functions defined in \cite{AS65} and $c_{1,2}$ are integration constants. From this, we can evaluate the second component. In the $y$ coordinates, it is given by
\begin{eqnarray}
 \Omega_{2}(y) &=& -\frac{1}{mc} \sqrt{\frac{2F}{c}} e^{i\frac{\pi}{4}} \left[ \frac{1}{2} y + \partial_{y} \right] \Omega_{1}(y) .
\end{eqnarray}
Then, using the recurrence relations for parabolic cylinder functions \cite{AS65}
\begin{eqnarray}
 \frac{1}{2} z U(a,z) + \partial_{z} U(a,z) &=& -\left(a+\frac{1}{2}\right)U(a+1,z), \\
\frac{1}{2} z U(a,z) - \partial_{z} U(a,z) &=& U(a-1,z) ,
\end{eqnarray}
we get that
\begin{eqnarray}
 \Omega_{2}(y) =  mc \sqrt{\frac{c}{2F}} e^{i\frac{3\pi}{4}}  c_{1} U(a+1,y) + \frac{1}{mc} \sqrt{\frac{2F}{c}} e^{-i\frac{\pi}{4}} c_{2} U(-a-1,-iy) .
\end{eqnarray}
Here and in the following , we define
\begin{eqnarray}
 U_{1}(a,y)&:=& U(a,y) ,\\ 
U_{2}(a,y) &:=&  U(-a,-iy), \\
 \tilde{U}_{1}(a,y)&:=&mc \sqrt{\frac{c}{2F}} e^{i\frac{3\pi}{4}} U(a+1,y) ,\\
 \tilde{U}_{2}(a,y) &:=& \frac{1}{mc} \sqrt{\frac{2F}{c}} e^{-i\frac{\pi}{4}} U(-a-1,y) ,
\end{eqnarray}
which allows us to write the general solution as
\begin{eqnarray}
  \psi_{1}(x) &=&  c_{1}U_{1}[a,y(x)]+c_{1}\tilde{U}_{1}[a,y(x)] + c_{2}U_{2}[a,y(x)] + c_{2}\tilde{U}_{2}[a,y(x)] , \\
i\psi_{2}(x) &=&   c_{1}U_{1}[a,y(x)]-c_{1}\tilde{U}_{1}[a,y(x)] + c_{2}U_{2}[a,y(x)] - c_{2}\tilde{U}_{2}[a,y(x)] ,
\end{eqnarray}
(the factors $1/\sqrt{2}$ were absorbed by redefining integration constants). Following the same procedure as in the preceding cases, the integration constants have to be determined from boundary conditions ((\ref{eq:init_cond_u}) and (\ref{eq:init_cond_v})) yielding $u$ and $v$ for $x \in (-R,R)$. The conditions in  (\ref{eq:bound_cond1_t}) and (\ref{eq:bound_cond2_t}) will serve to find $u^{>}$ and $v^{>}$ for $x \in (R, \infty)$, while (\ref{eq:bound_cond3_t}) and (\ref{eq:bound_cond4_t}) will serve to find $u^{<}$ and $v^{<}$ for $x \in (-\infty,-R)$.

It is now possible to evaluate the integration constants such that the boundary conditions  (\ref{eq:init_cond_u}) and (\ref{eq:init_cond_v}) are obeyed. Letting $y_{0} := e^{-i\frac{\pi}{4}}\sqrt{\frac{2c}{F}} \left( \frac{E}{c} \right)$ be the value where $x=0$, we find the coefficients
\begin{eqnarray}
 c_{1} &=& - \frac{1}{2} \frac{U_{2}(a,y_{0}) - \tilde{U}_{2}(a,y_{0})}{\tilde{U}_{2}(a,y_{0})U_{1}(a,y_{0}) - U_{2}(a,y_{0})\tilde{U}_{1}(a,y_{0})} ,\\
 c_{2} &=&  \frac{1}{2}\frac{U_{1}(a,y_{0}) - \tilde{U}_{1}(a,y_{0})}{\tilde{U}_{2}(a,y_{0})U_{1}(a,y_{0}) - U_{2}(a,y_{0})\tilde{U}_{1}(a,y_{0})} ,
\end{eqnarray}
and
\begin{eqnarray}
 c'_{1} &=& \frac{i}{2} \frac{U_{2}(a,y_{0}) + \tilde{U}_{2}(a,y_{0})}{\tilde{U}_{2}(a,y_{0})U_{1}(a,y_{0}) - U_{2}(a,y_{0})\tilde{U}_{1}(a,y_{0})} ,\\
 c'_{2} &=& -\frac{i}{2} \frac{U_{1}(a,y_{0}) + \tilde{U}_{1}(a,y_{0})}{\tilde{U}_{2}(a,y_{0})U_{1}(a,y_{0}) - U_{2}(a,y_{0})\tilde{U}_{1}(a,y_{0})} ,
\end{eqnarray}
for the first ($u$ has integration constants $c_{1,2}$) and second ($v$ has integration constants $c'_{1,2}$) boundary conditions respectively. The next step is to compute the solution for $x \in (-\infty,-R)$ and $x \in (R, \infty)$.

The general solutions for $x \in (R, \infty)$ can be written as
\begin{eqnarray}
 u^{>}_{1}(x) &=&   b_{1}U_{1}[a,y(x)]+b_{1}\tilde{U}_{1}[a,y(x)] + b_{2}U_{2}[a,y(x)] + b_{2}\tilde{U}_{2}[a,y(x)] , \\
iu^{>}_{2}(x) &=&  b_{1}U_{1}[a,y(x)]-b_{1}\tilde{U}_{1}[a,y(x)] + b_{2}U_{2}[a,y(x)] - b_{2}\tilde{U}_{2}[a,y(x)],  \\
 v^{>}_{1}(x) &=&   b'_{1}U_{1}[a,y(x)]+b'_{1}\tilde{U}_{1}[a,y(x)] + b'_{2}U_{2}[a,y(x)] + b'_{2}\tilde{U}_{2}[a,y(x)] , \\
iv^{>}_{2}(x) &=&  b'_{1}U_{1}[a,y(x)]-b'_{1}\tilde{U}_{1}[a,y(x)] + b'_{2}U_{2}[a,y(x)] - b'_{2}\tilde{U}_{2}[a,y(x)] , 
\end{eqnarray}
while for $x \in (-\infty,-R)$, we have
\begin{eqnarray}
 u^{<}_{1}(x) &=&   a_{1}U_{1}[a,y(x)]+a_{1}\tilde{U}_{1}[a,y(x)] + a_{2}U_{2}[a,y(x)] + a_{2}\tilde{U}_{2}[a,y(x)] , \\
iu^{<}_{2}(x) &=&  a_{1}U_{1}[a,y(x)]-a_{1}\tilde{U}_{1}[a,y(x)] + a_{2}U_{2}[a,y(x)] - a_{2}\tilde{U}_{2}[a,y(x)] , \\
 v^{<}_{1}(x) &=&   a'_{1}U_{1}[a,y(x)]+a'_{1}\tilde{U}_{1}[a,y(x)] + a'_{2}U_{2}[a,y(x)] + a'_{2}\tilde{U}_{2}[a,y(x)] , \\
iv^{<}_{2}(x) &=&  a'_{1}U_{1}[a,y(x)]-a'_{1}\tilde{U}_{1}[a,y(x)] + a'_{2}U_{2}[a,y(x)] - a'_{2}\tilde{U}_{2}[a,y(x)],  
\end{eqnarray}
where the new coefficients $a_{1,2},a'_{1,2},b_{1,2},b'_{1,2}$ can be determined from the boundary conditions in  (\ref{eq:bound_cond1_t}) to (\ref{eq:bound_cond4_t}). First, let us define the following constants:
\begin{eqnarray}
G_{u}^{\pm} &:=& \left[1+ \frac{g^{2}}{4c^{2}} \right]^{-1} \left[ \left( 1- \frac{g^{2}}{4c^{2}} \right) u_{2,1}(R) \pm \frac{g}{c} u_{1,2}(R)  \right], \\
G_{v}^{\pm} &:=& \left[1+ \frac{g^{2}}{4c^{2}} \right]^{-1} \left[ \left( 1- \frac{g^{2}}{4c^{2}} \right) v_{2,1}(R) \pm \frac{g}{c} v_{1,2}(R)  \right] ,\\
H_{u}^{\pm} &:=& \left[1+ \frac{g^{2}}{4c^{2}} \right]^{-1} \left[ \left( 1- \frac{g^{2}}{4c^{2}} \right) u_{1,2}(-R) \pm \frac{g}{c} u_{2,1}(-R)  \right], \\
H_{v}^{\pm} &:=& \left[1+ \frac{g^{2}}{4c^{2}} \right]^{-1} \left[ \left( 1- \frac{g^{2}}{4c^{2}} \right) v_{1,2}(-R) \pm \frac{g}{c} v_{2,1}(-R)  \right] .
\end{eqnarray}
The boundary conditions at $x=R$ now become
\begin{eqnarray}
u^{>}_{1}(R) &=& G_{u}^{-} \;\; , \;\;
u^{>}_{2}(R) = G_{u}^{+} ,\\
v^{>}_{1}(R) &=& G_{v}^{-} \;\; , \;\;
v^{>}_{2}(R) = G_{v}^{+} ,
\end{eqnarray}
while at $x=-R$, they are given by
\begin{eqnarray}
u^{<}_{1}(-R) &=& H_{u}^{+} \;\; , \;\;
u^{<}_{2}(-R) = H_{u}^{-} ,\\
v^{<}_{1}(-R) &=& H_{v}^{+} \;\; , \;\;
v^{<}_{2}(-R) = H_{v}^{-} .
\end{eqnarray}
It is now possible to solve for $a_{1,2},a'_{1,2},b_{1,2},b'_{1,2}$ from these equations. We get
\begin{eqnarray}
\label{eq:coeff_sol1}
b_{2} &=& \frac{1}{2} \frac{iG_{u}^{+} U^{+}_{1}[a,y(R)]  - G_{u}^{-} U^{-}_{1}[a,y(R)]}{\tilde{U}_{1}[a,y(R)]U_{2}[a,y(R)] - U_{1}[a,y(R)]\tilde{U}_{2}[a,y(R)]} ,\\
\label{eq:coeff_sol2}
b'_{2} &=&  \frac{1}{2} \frac{iG_{v}^{+} U^{+}_{1}[a,y(R)]  - G_{v}^{-} U^{-}_{1}[a,y(R)]}{\tilde{U}_{1}[a,y(R)]U_{2}[a,y(R)] - U_{1}[a,y(R)]\tilde{U}_{2}[a,y(R)]},\\
\label{eq:coeff_sol3} 
a_{1} &=&  -\frac{1}{2} \frac{iH_{u}^{-} U^{+}_{2}[a,y(-R)]  - H_{u}^{+} U^{-}_{2}[a,y(-R)] }{\tilde{U}_{1}[a,y(-R)]U_{2}[a,y(-R)] - U_{1}[a,y(-R)]\tilde{U}_{2}[a,y(-R)]} ,\\
\label{eq:coeff_sol4}
a'_{1} &=&  -\frac{1}{2} \frac{iH_{v}^{-} U^{+}_{2}[a,y(-R)]  - H_{v}^{+} U^{-}_{2}[a,y(-R)]}{\tilde{U}_{1}[a,y(-R)]U_{2}[a,y(-R)] - U_{1}[a,y(-R)]\tilde{U}_{2}[a,y(-R)]} ,
\end{eqnarray}
where $U^{\pm}_{1,2}:=U_{1,2} \pm \tilde{U}_{1,2}$ and similar expressions for $b_{1},b'_{1},a_{2},a'_{2}$. However, as will be seen in the next section, the latter are not useful for the spectral density calculation. 

The preceding calculation allowed us to evaluate the solutions $u^{\pm}$ and $v^{\pm}$. Remember that these solutions are defined on $x>0$ and $x<0$ (see Figure \ref{fig:dom_sol} for the domain of each function). Moreover, these solutions still obey the boundary conditions of the WTK method, that is  (\ref{eq:init_cond_u}) and (\ref{eq:init_cond_v}), and can be written as
\begin{eqnarray}
 u^{+}(x) &=& \left\{
\begin{matrix}
 u(x) & \mbox{for} \; x \in [0,R) \\
u^{>}(x) & \mbox{for} \; x \in (R, \infty)
\end{matrix} \right. 
,
 u^{-}(x) = \left\{
\begin{matrix}
 u(x) & \mbox{for} \; x \in (-R,0] \\
u^{<}(x) & \mbox{for} \; x \in (-\infty,-R)
\end{matrix} \right. ,
\\
 v^{+}(x) &=& \left\{
\begin{matrix}
 v(x) & \mbox{for} \; x \in [0,R) \\
v^{>}(x) & \mbox{for} \; x \in (R,\infty)
\end{matrix} \right.
,
 v^{-}(x) = \left\{
\begin{matrix}
 v(x) & \mbox{for} \; x \in (-R,0] \\
v^{<}(x) & \mbox{for} \; x \in (-\infty,-R)
\end{matrix} \right. .
\end{eqnarray}
We now have all the ingredients to compute the Titchmarsh $m$-functions and the spectral density.

\subsection{Evaluation of the Weyl-Titchmarsh $m$-functions}

The $m$-functions are constructed such that the solutions $\psi^{\pm} = u^{\pm} + m^{\pm}(E) v^{\pm}$ are finite when $x \rightarrow \pm \infty$ (with the energy obeying $\Im(E) >0 $). This can be evaluated by looking at the asymptotic expansion of the solutions. Asymptotically, in the limit $z \gg |a|$, the parabolic cylinder functions have the following behavior \cite{AS65}:
\begin{eqnarray}
U(a,z) &\sim &  e^{-\frac{1}{4} z^{2}} z^{-a-\frac{1}{2}} \;\; \mbox{for} \;\; |\arg(z) |< \frac{3}{4}\pi ,\\
U(a,z) &\sim &  e^{-\frac{1}{4}z^{2}} z^{-a-\frac{1}{2}} \pm i\frac{\sqrt{2\pi}}{\Gamma \left( \frac{1}{2} - a \right)} e^{\mp i \pi a} e^{\frac{1}{4}z^{2}} z^{a-\frac{1}{2}} \nonumber \\&& \mbox{for} \;\; \frac{1}{4}\pi < \pm \arg(z) < \frac{5}{4}\pi ,
\end{eqnarray}
where $\Gamma$ is the usual Gamma function. From these relations and the fact that $ y = e^{-i\frac{\pi}{4}}\sqrt{\frac{2c}{F}} \left( \frac{E+Fx}{c} \right)$, we find that for the solution $\psi^{+}$ to be in $L^{2}(0,\infty) \otimes \mathbb{C}^{2}$ when $\Im (E) >0$, it should be 
\begin{eqnarray}
\label{eq:as_sol1}
 \lim_{x \rightarrow + \infty}\psi^{+}_{1}(x) & =&  \lim_{x \rightarrow + \infty} (b_{1}+m^{+}b_{1}') \left\{ U_{1}[a,y(x)] + \tilde{U}_{1}[a,y(x)] \right\}   = 0 ,\\
\label{eq:as_sol3}
 \lim_{x \rightarrow + \infty} \psi^{+}_{2}(x)& =& -i \lim_{x \rightarrow + \infty} (b_{1}+m^{+}b_{1}') \left\{ U_{1}[a,y(x)] + \tilde{U}_{1}[a,y(x)] \right\} = 0,
\end{eqnarray}
as $x \rightarrow \infty$, whereas for the solution $\psi^{-}$ to be in $L^{2}(-\infty,0) \otimes \mathbb{C}^{2}$ when $\Im (E) >0$, it should be
\begin{eqnarray}
\label{eq:as_sol2}
 \lim_{x \rightarrow - \infty}\psi^{-}_{1}(x) & =&   \lim_{x \rightarrow - \infty} (a_{2}+m^{-}a_{2}') \left\{ U_{2}[a,y(x)] + \tilde{U}_{2}[a,y(x)] \right\}  = 0,\\
\label{eq:as_sol4}
 \lim_{x \rightarrow - \infty}\psi^{-}_{2}(x) & =&   -i \lim_{x \rightarrow - \infty}(a_{2}+m^{-}a_{2}') \left\{ U_{2}[a,y(x)] + \tilde{U}_{2}[a,y(x)] \right\} = 0,
\end{eqnarray}
as $x \rightarrow -\infty$. The other terms in the solutions would make the limits diverge in the last four expressions. Thus, the $m$-functions have to be chosen such that a form similar to  (\ref{eq:as_sol1}),(\ref{eq:as_sol3}),(\ref{eq:as_sol2}) and (\ref{eq:as_sol4}) is recovered, which is achieved by choosing
\begin{eqnarray}
\label{eq:m_func_ab}
 m^{+}(E) = -\frac{b_{2}}{b'_{2}} \;\;\mbox{and}\;\; m^{-}(E) = -\frac{a_{1}}{a'_{1}}.
\end{eqnarray}
These relations are very important because they relate the $m$-functions to the explicit solution allowing us to evaluate the spectral density function. 

Substituting  (\ref{eq:coeff_sol1}) to (\ref{eq:coeff_sol4}) in (\ref{eq:m_func_ab}), we get finally
\begin{eqnarray}
\label{eq:m_plus}
m^{+}(E)&=&  \frac{G_{u}^{-} U^{-}_{1}[a,y(R)] -iG_{u}^{+} U^{+}_{1}[a,y(R)]}{iG_{v}^{+} U^{+}_{1}[a,y(R)]  - G_{v}^{-} U^{-}_{1}[a,y(R)] } ,\\
\label{eq:m_minus} 
m^{-}(E)&=&  \frac{H_{u}^{+} U^{-}_{2}[a,y(-R)] -iH_{u}^{-} U^{+}_{2}[a,y(-R)] }{iH_{v}^{-} U^{+}_{2}[a,y(-R)] - H_{v}^{+} U^{-}_{2}[a,y(-R)] } .
\end{eqnarray}

\subsection{Spectral density and physical interpretation}

Once the Weyl-Titchmarsh functions are known, it is straightforward to compute the spectral density using  (\ref{eq:spec_trace}), (\ref{eq:m_plus}) and (\ref{eq:m_minus}). It is a complicated functions of the interatomic distance $R$, the potential strength $g$ and the electric field strength $F$. Because the potential is not bounded, it is expected that the spectrum will be a continuum on $E \in \mathbb{R}$. Bound states on the other hand should become resonances for which the width is related to the decay time of these quasibound states. To analyze these assumptions, the spectral density is plotted in the following and compared to the case without electric field. The position of poles of $\rho$ in the complex energy plane is also investigated. Finally, the ``plunging'' of the ground state resonance in the Dirac sea is discussed. Notice here that this section is included to show some qualitative features of the model; a more rigorous treatment would be required to understand all of these results and is presently under study.

\subsubsection{Dependence on field strength and interatomic distance}

\begin{figure}
\subfloat[]{\includegraphics[width=3.0in]{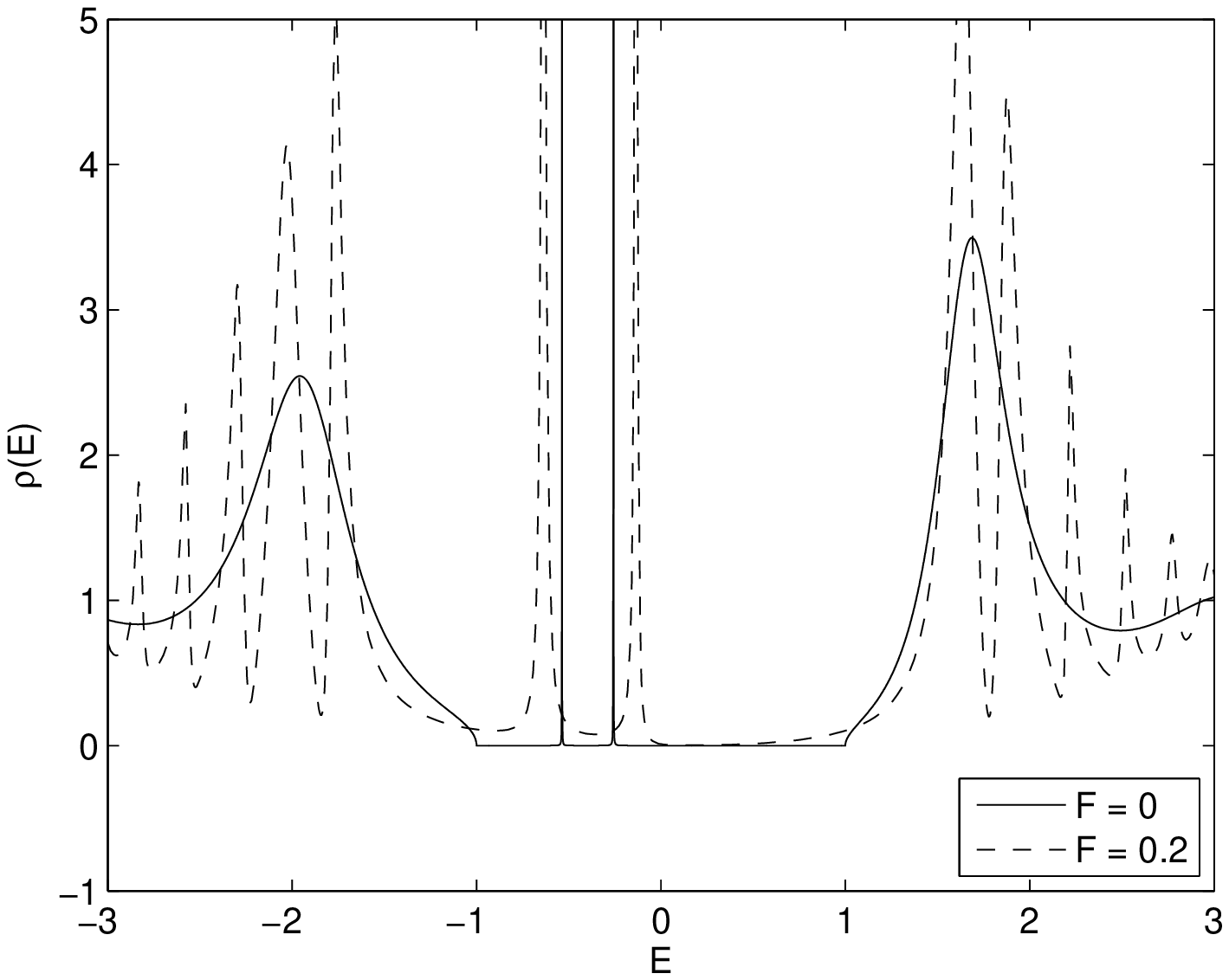}}
\subfloat[]{\includegraphics[width=3.0in]{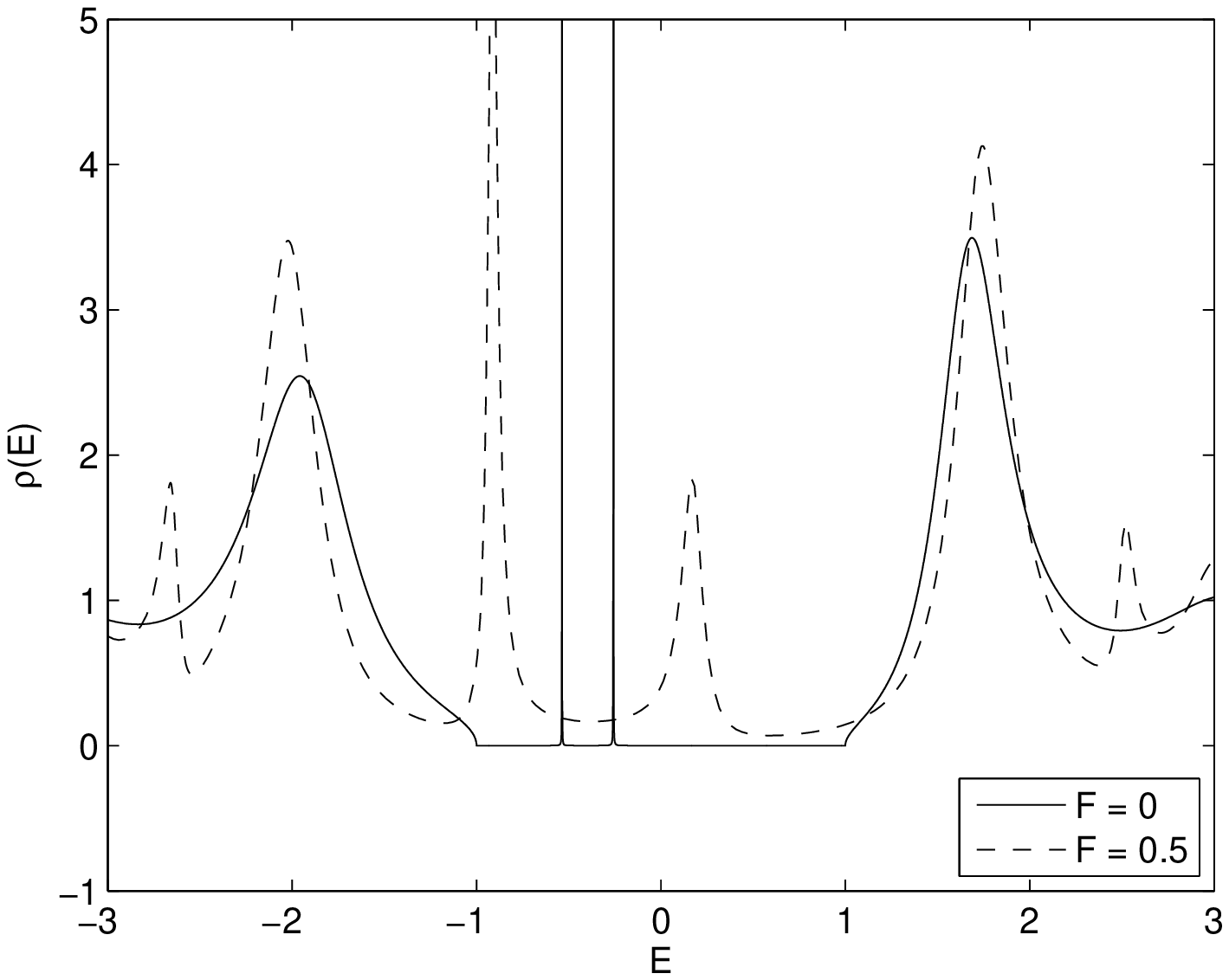}} \\
\subfloat[]{\includegraphics[width=3.0in]{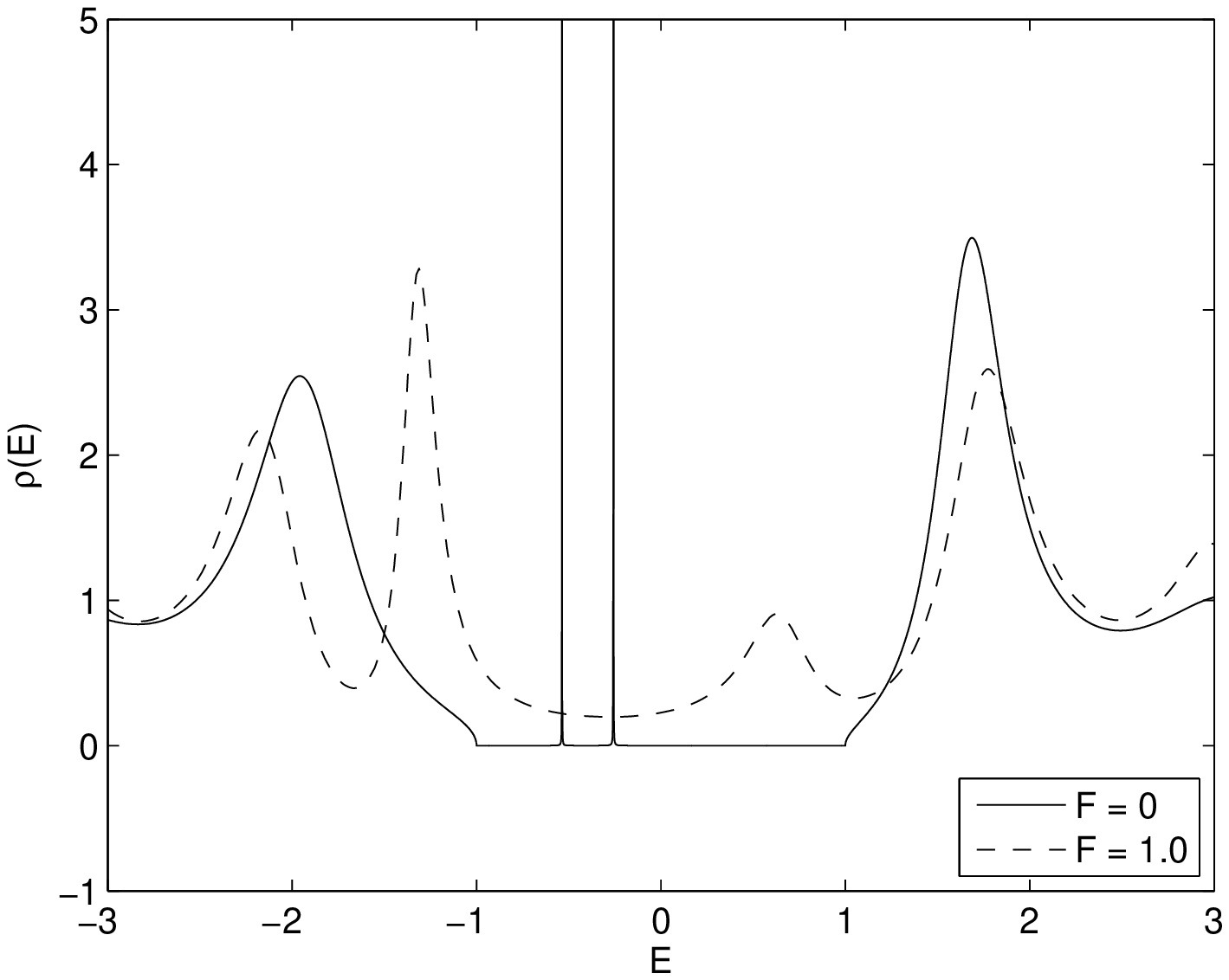}} 
\subfloat[]{\includegraphics[width=3.0in]{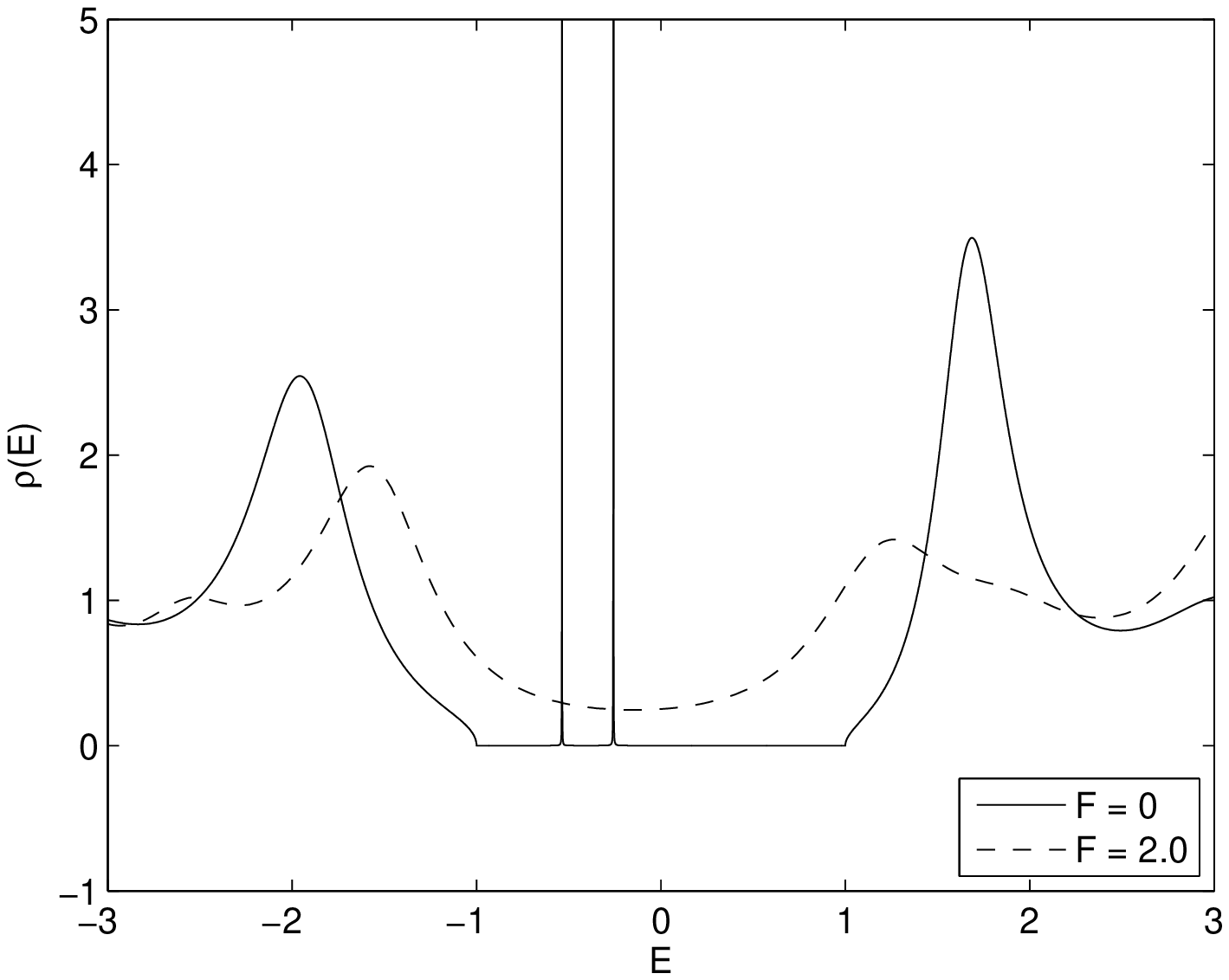}}
\caption{Spectral density for a two Dirac delta potential for varying values of the electric field strength $F$. The value of the potential strength is fixed to $g=3.0$, the value of the interatomic distance is $R=1.0$ while the the electric field is (a) $F=0.2$, (b) $F=0.5$, (c) $F=1.0$ and (d) $F=2.0$.}
\label{fig:rho_2_del_with_F}
\end{figure}

In Figure \ref{fig:rho_2_del_with_F}, the spectral density for different values of the field strength is shown and compared to the case without electric field. As discussed previously, the system has two bound states when $F=0$ (if the delta potential strength $g$ lies in the interval $-4mc^{3}R + 2c \sqrt{4m^{2}c^{4}R^{2}+1} \leq g<4mc^{3}R + 2c \sqrt{4m^{2}c^{4}R^{2}+1}$). As the electric field is turned on, the bound states become Stark-shifted unstable states: the ground state and first excited states are shifted in opposite direction while their width increases. This behavior is very similar to the non-relativistic results \cite{PhysRevA.69.053409}.

It can be understood mathematically as the occurrence of resonances, that is poles on the unphysical sheet of the analytically continued resolvent operator \cite{reed1972methods} (or Green's function) given by $(H-E)^{-1}$ (assuming of course that it admits an analytic continuation to the lower half complex plane in $E$). Recall that the resolvent operator has the same analytical structure in the complex energy $E$ plane as the ``trace spectral density'' because the former is given by \cite{nla.cat-vn2169453}
\begin{eqnarray}
 G_{E}(x,y) = \left\{
\begin{array}{l}
\frac{\psi^{+}(x)\psi^{-}(y)}{m^{+}(E) - m^{-}(E)}, \; \mbox{for}\; x\leq y \\
\frac{\psi^{+}(y)\psi^{-}(x)}{m^{+}(E) - m^{-}(E)}, \; \mbox{for}\; x > y 
\end{array}
\right.
\end{eqnarray}
where $\psi^{\pm}(x) = u^{\pm}(x) + m^{\pm}(E) v^{\pm}(x)$. The solutions $u^{\pm}$ and $v^{\pm}$ have no poles in $E$, so the analytical structure in the energy plane would come from the terms:
\begin{eqnarray}
\frac{1}{m^{+}(E) - m^{-}(E)} \;\; \mbox{and} \;\; \frac{m^{+}(E)m^{-}(E)}{m^{+}(E) - m^{-}(E)}
\end{eqnarray}
which are included in $\rho$. When there is no electric field, there are poles lying on $E \in \R$ (bound states) and are solutions to \eqref{eq:pole_pos}. When the electric field is nonzero, these poles move to the unphysical sheet and becomes eigensolutions to a non-hermitian operator \cite{springerlink:10.1007/BF00398170} which allows complex valued eigenvalues. The effect of these poles shows up in the presence of peaks in the spectral density (if the poles are close enough to the real axis). 

The continua also show the main features of the non-relativistic results: as the field strength decreases, the oscillation increases and average out to the zero-field case in the limit $F \rightarrow 0$. For each bump of the continua corresponds a pole in the complex energy plane. It would be interesting to study if the branch cuts on $(-\infty, -mc^{2}]$ and $[mc^{2},\infty)$ when $F=0$ degenerates into these poles as the electric field is turned on. Physically, these resonances were interpreted in \cite{PhysRevA.69.053409} as an example of Ramsauer-Townsend resonances. The figure also shows that there are two frequencies of oscillation. The first one corresponds to the backscattering of the electron by the two delta potential wells (inducing a smaller frequency of oscillation and also existing in the case without electric field) while the second one is associated to the scattering on the whole system, including the electric potential (inducing a larger frequency for small enough $F$)\cite{PhysRevA.69.053409}.

\begin{figure}
\subfloat[]{\includegraphics[width=3.0in]{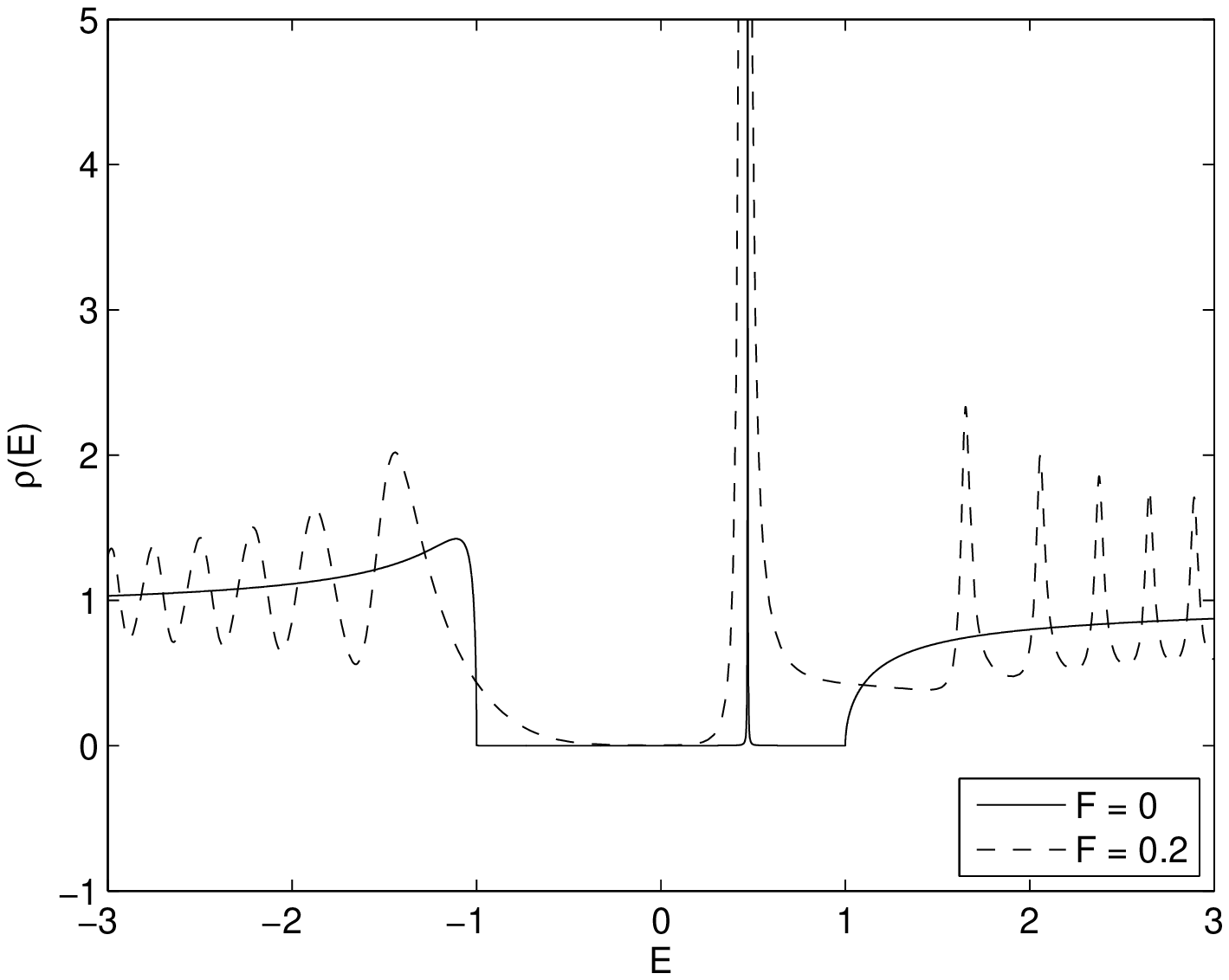}}
\subfloat[]{\includegraphics[width=3.0in]{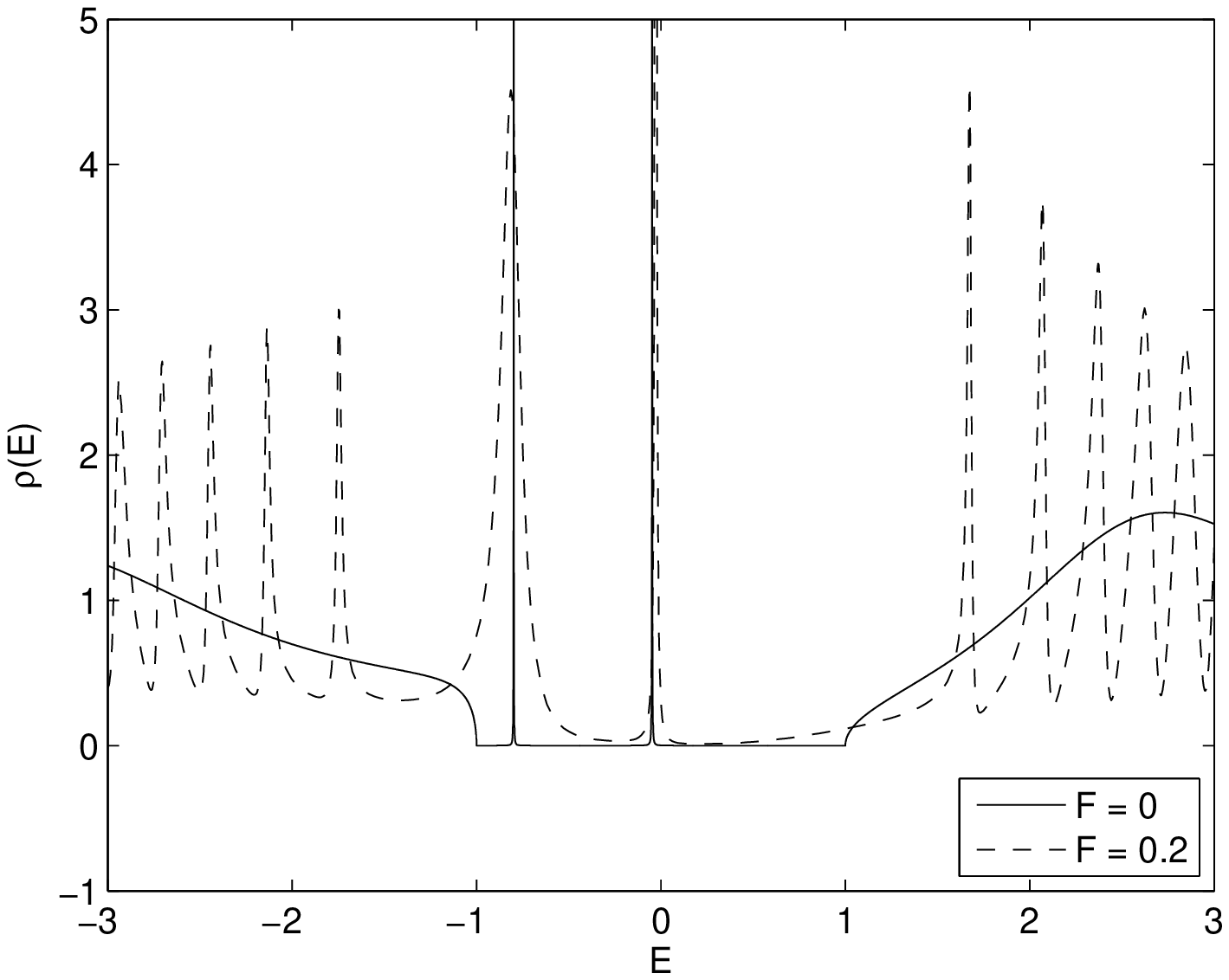}} \\
\subfloat[]{\includegraphics[width=3.0in]{with_field_g3_F02_R1}} 
\subfloat[]{\includegraphics[width=3.0in]{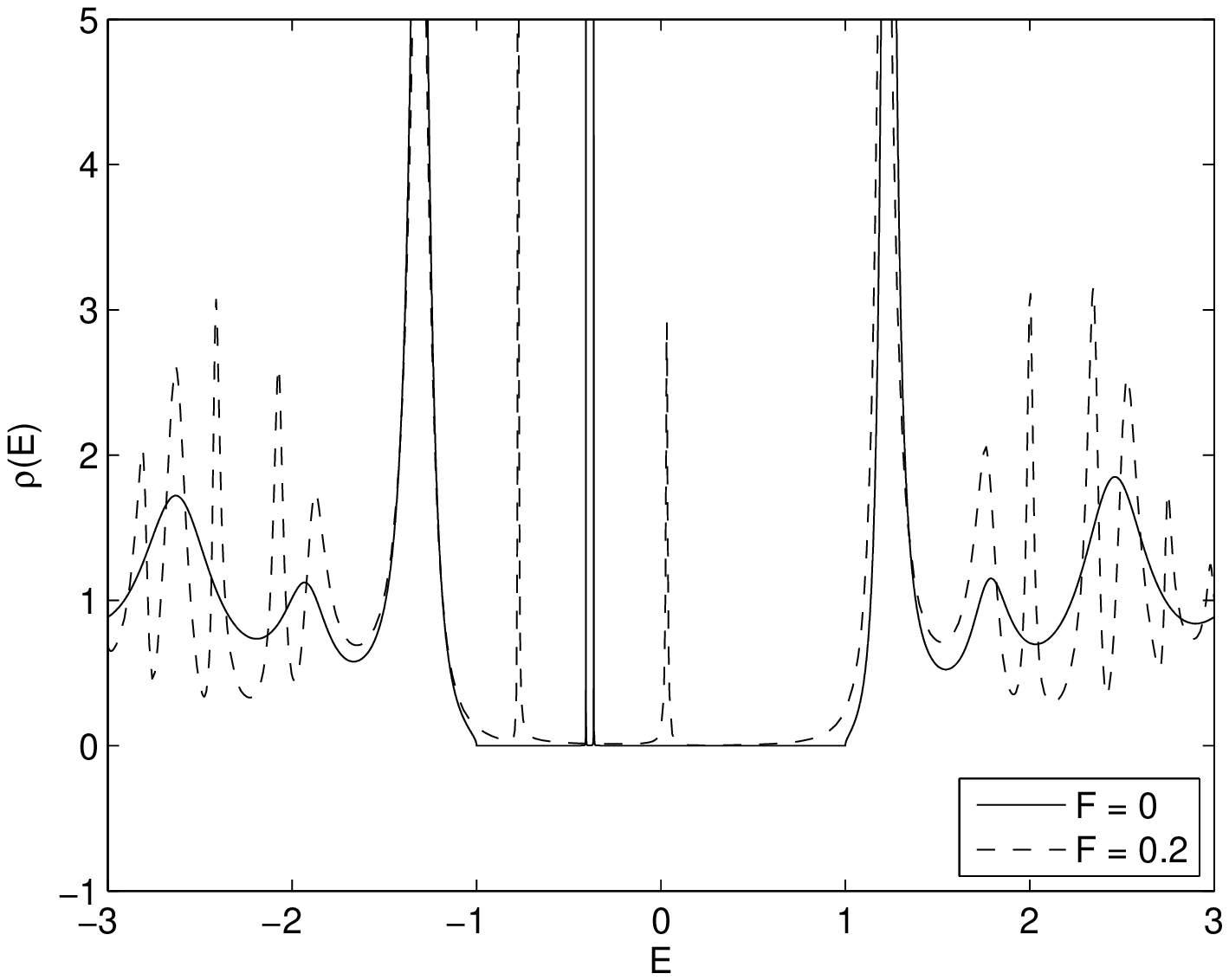}}
\caption{Spectral density for a two Dirac delta potential and an electric field of strength $F=0.2$, for varying values of the interatomic distance $R$. The value of the potential strength is fixed to $g=3.0$ while the the interatomic distance is (a) $R=0.1$, (b) $R=0.5$, (c) $R=1.0$ and (d) $R=2.0$.}
\label{fig:rho_2_del_with_R}
\end{figure}

In Figure \ref{fig:rho_2_del_with_R}, the spectral density for various interatomic distances and fixed value of electric field strength ($F=0.2$) is presented. It is well-known that as $R \rightarrow \infty$ in the zero-field case, the two bound states becomes quasi-degenerate \cite{PhysRevA.69.053409}. This is because the two potential wells become independent as the typical tunneling probability between them $P_{\rm tunnel} \rightarrow 0$ as $R \rightarrow \infty$. In the presence of the electric field, the two bound states are shifted and do not become degenerate, as shown in the picture. As the interatomic distance is increased, their energy shift becomes more important (the symmetric ground state is pushed to lower energy while the anti-symmetric excited state is pulled to higher energy) and their widths decrease. The latter is due to a larger tunneling barrier for the electron as $R$ becomes larger. All of this is very similar to what one would expect from the non-relativistic analysis \cite{PhysRevA.69.053409}. It also suggests a new process to probe the negative energy continuum: by taking $R$ large enough for a given value of $F$, the real part of the ground state resonance could go into a region where it overlaps with negative energy states. We will see in Section \ref{sec:dirac_sea} however that this interpretation is slightly too simplistic as avoided crossings may occur depending on the value of $g$. 

\subsubsection{``Plunging'' deeper in the Dirac sea}
\label{sec:dirac_sea}

\begin{figure}
\subfloat[]{\includegraphics[width=3.0in]{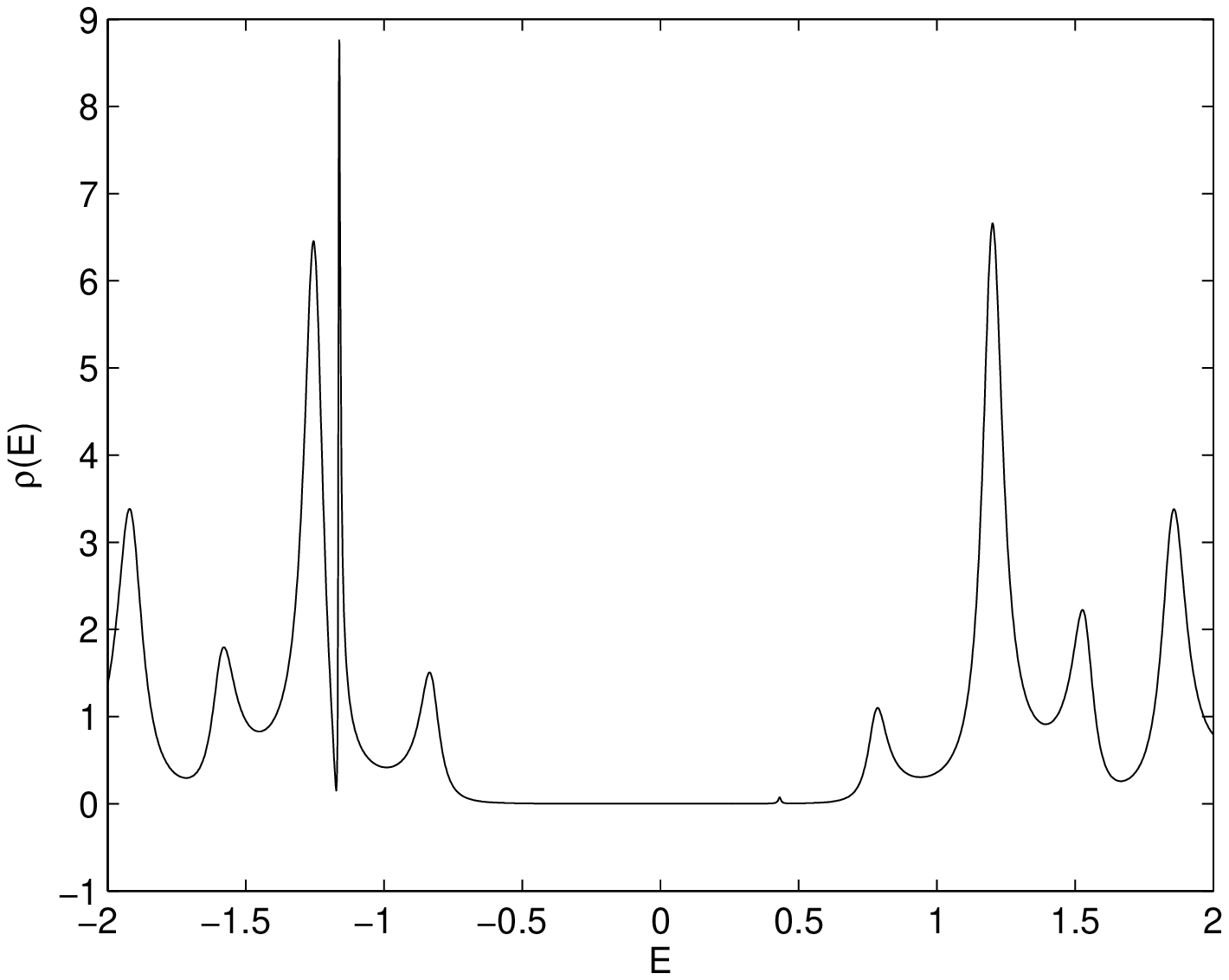}}
\subfloat[]{\includegraphics[width=3.0in]{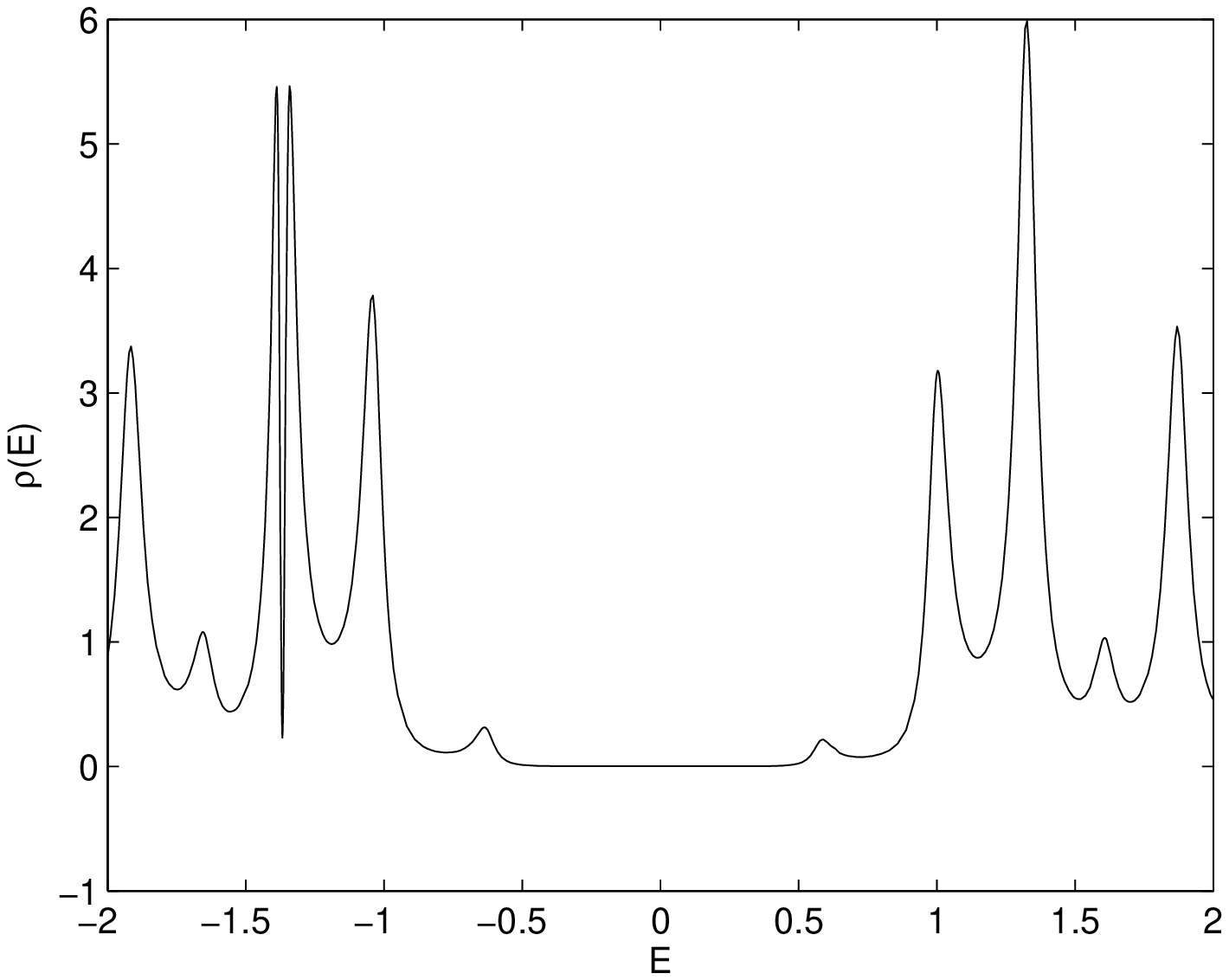}} \\
\subfloat[]{\includegraphics[width=3.0in]{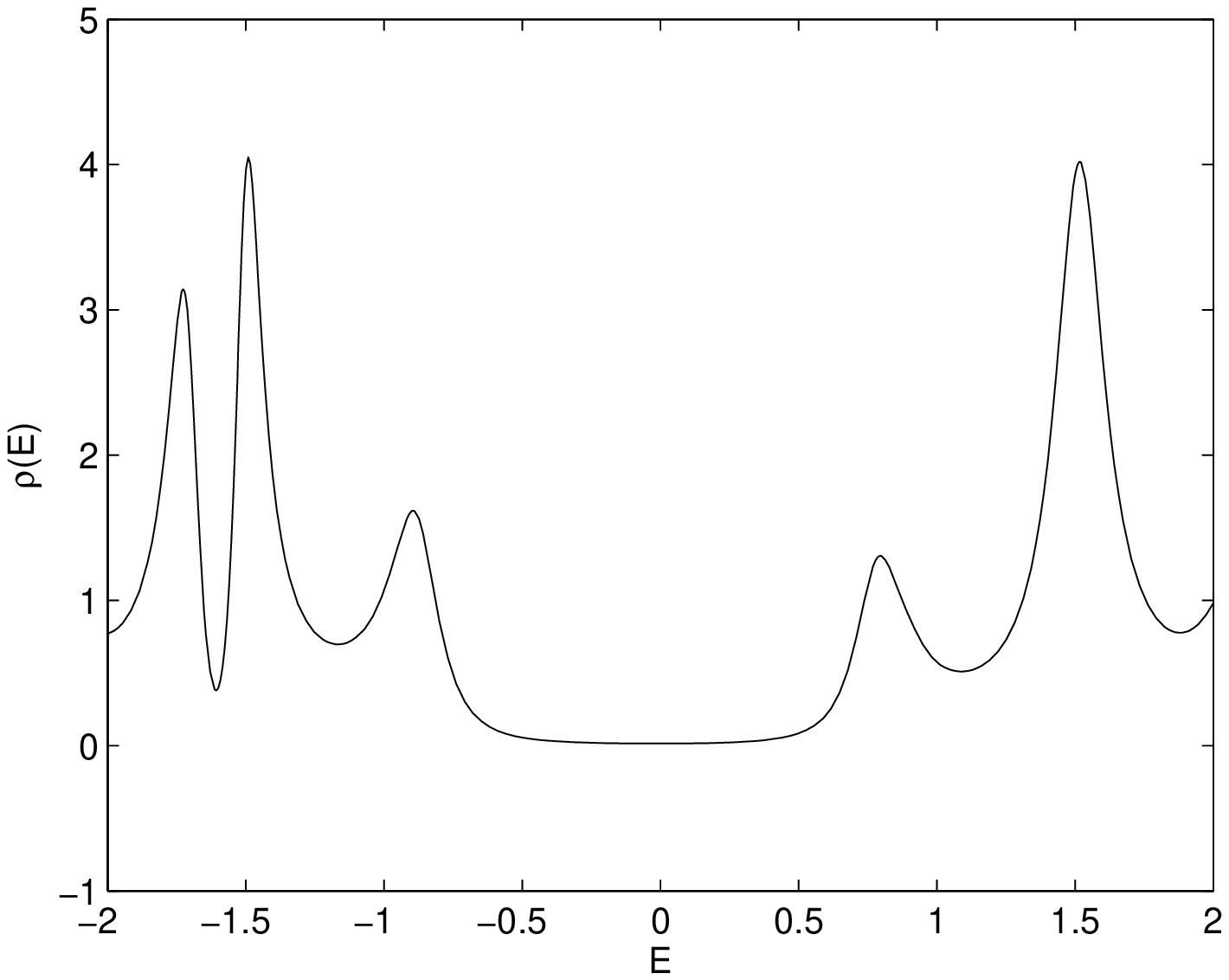}}
\subfloat[]{\includegraphics[width=3.0in]{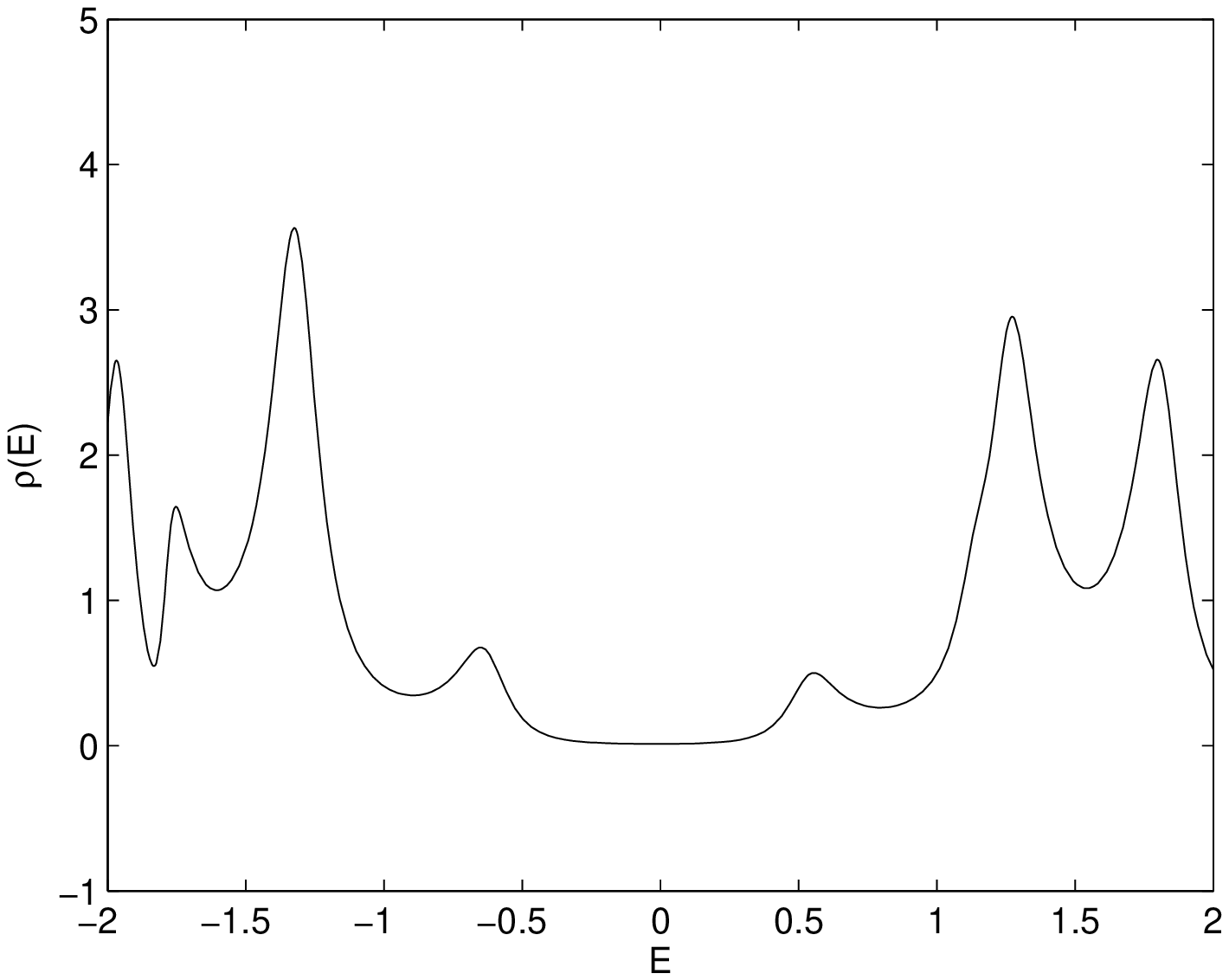}}
\caption{Spectral density for a two Dirac delta potential. The value of the potential strength is fixed to $g=3.0$ while the other parameters are (a) $F=0.2$ and $R=4.0$, (b) $F=0.2$ and $R=5.0$, (c) $F=0.5$ and $R=2.5$, (d) $F=0.5$ and $R=3.0$. There is a peak that goes into the negative energy continuum as $R$ is increased, it is the fourth one from the left in (a) and (b) and the second one in (c) and (d).}
\label{fig:rho_dirac_sea}
\end{figure}

For a given value of the electric strength, there is a critical value for the interatomic distance $R$ for which the resonance energy $E^{F}_{\rm ground} = \Re(E^{F}_{\rm ground}) + i\Im(E^{F}_{\rm ground})$ associated to the ground state (that is $\lim_{F \rightarrow 0}E^{F}_{\rm ground} = E^{0}_{\rm ground}$ where $E^{0}_{\rm ground}$ is a solution to \eqref{eq:pole_pos}) approaches the negative energy continuum of the free operator, that is $ \Re(E^{F}_{\rm ground}) \gtrsim -mc^{2}$. As the interatomic distance is increased further, one would expect that the resonance would go down below $-mc^{2}$ where most of the negative energy continuum states belongs. This phenomenon is hinted by the results obtained for the spectral density at large interatomic distance. A few examples of this are shown in Figure \ref{fig:rho_dirac_sea} where the spectral density for two electric strengths ($F=0.2$ and $F=0.5$) and two interatomic distances are presented. As seen in these pictures, the ground state resonance seems to cross the Ramsauer-Townsend resonances and seems to keep going to more negative energy and larger width as the interatomic distance is increased. 

\begin{figure}
\subfloat[]{\includegraphics[width=3.0in]{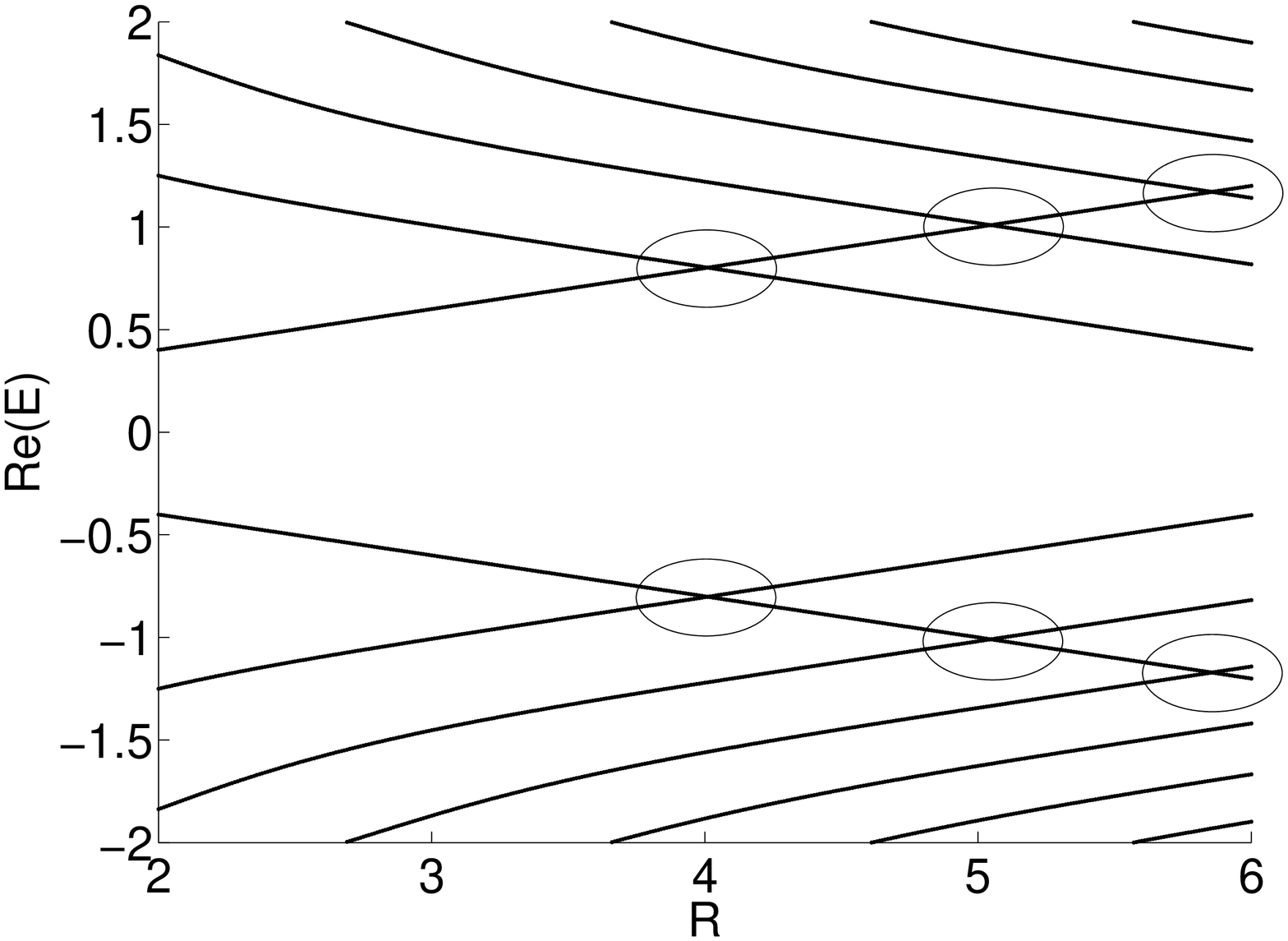}}
\subfloat[]{\includegraphics[width=3.0in]{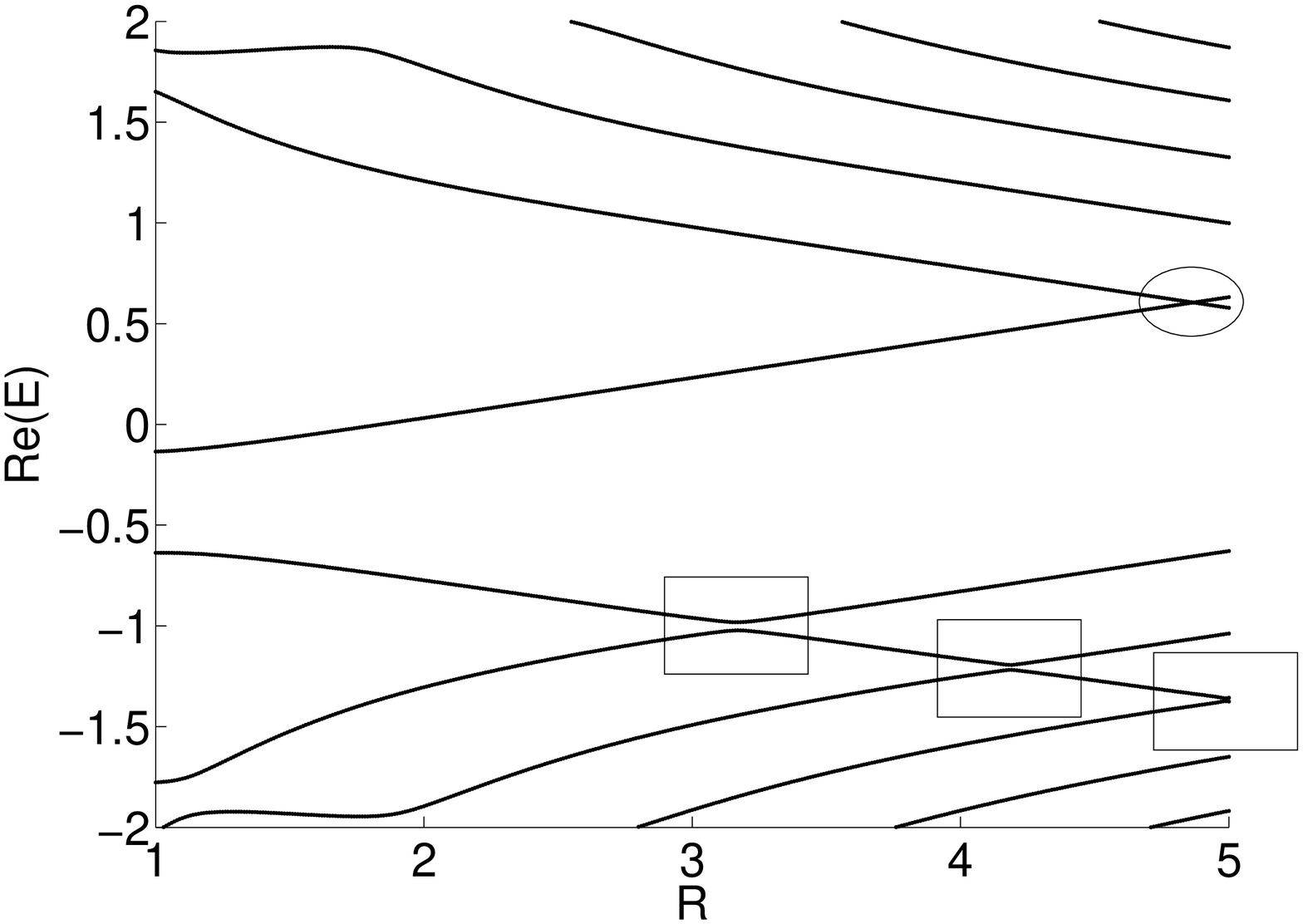}} \\
\subfloat[]{\includegraphics[width=3.0in]{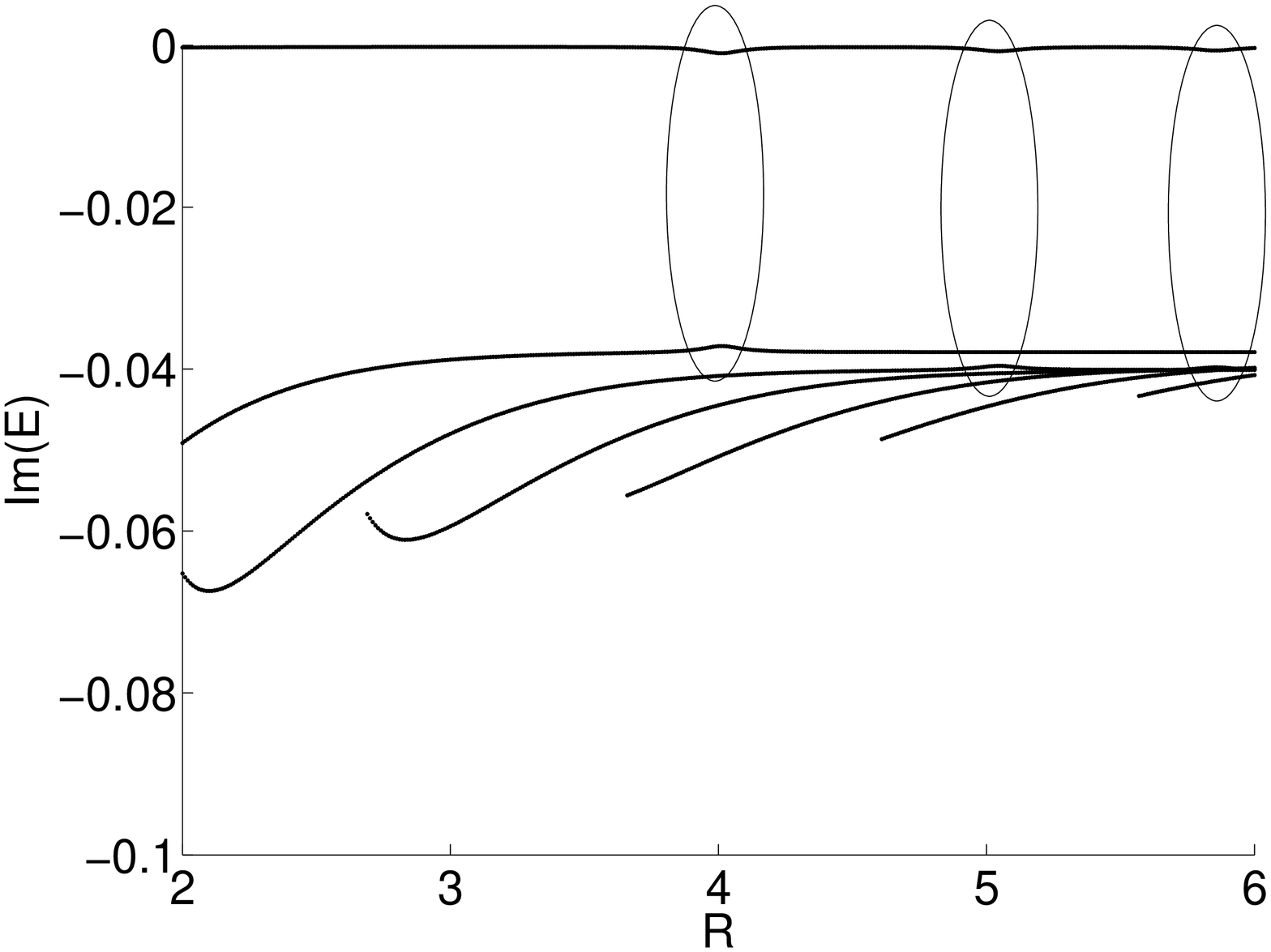}}
\subfloat[]{\includegraphics[width=3.0in]{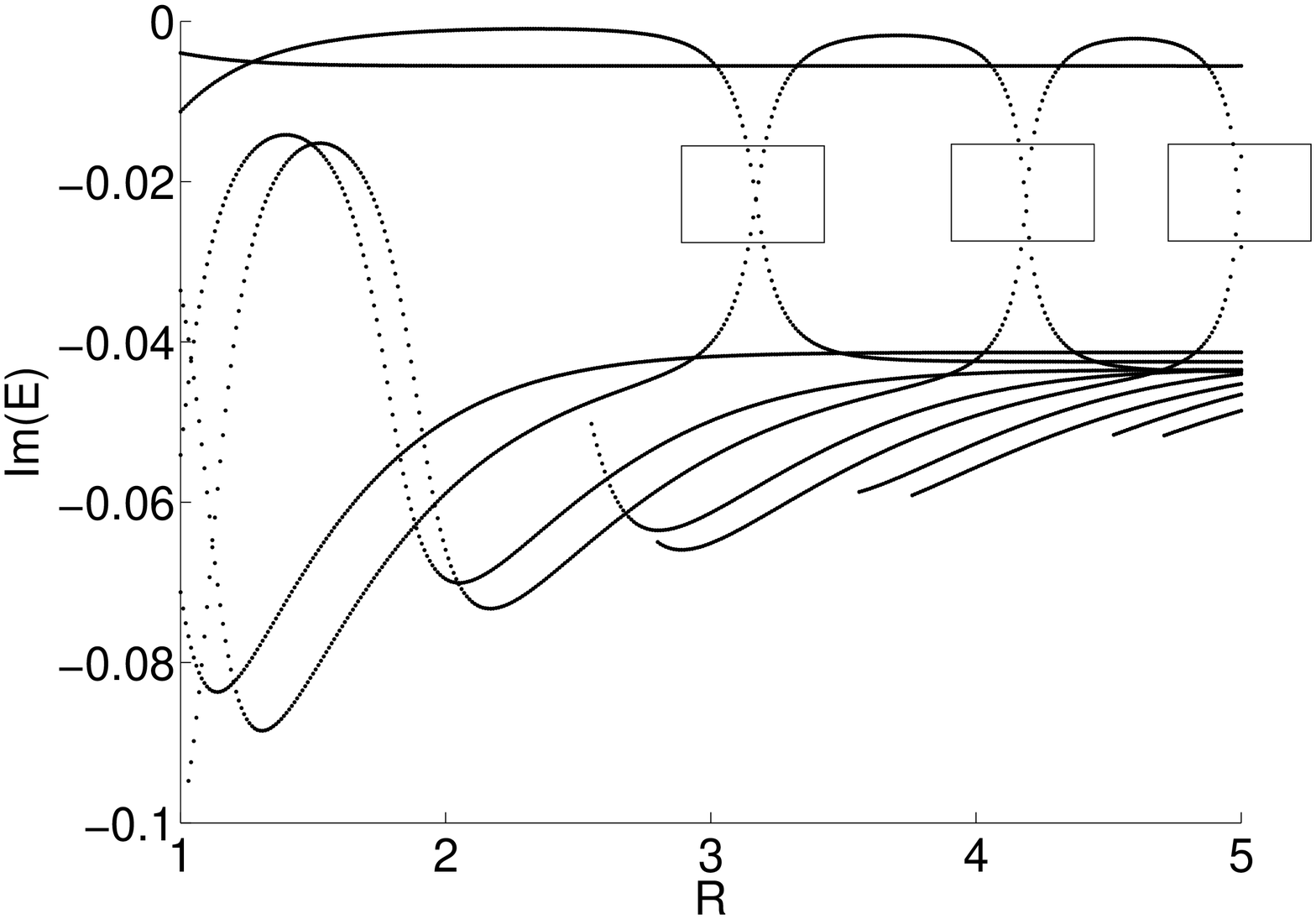}} \\
\caption{Real and imaginary part of the poles in the complex energy plane for different values of $R$ and for $F=0.2$. In  (a) and (c), we have $g=2.0$ while for (b) and (d), we have $g=3.0$. The avoided crossings are surrounded by rectangles while the crossings are encircled.}
\label{fig:pole_position}
\end{figure}

\begin{figure}
\subfloat[]{\includegraphics[width=3.0in]{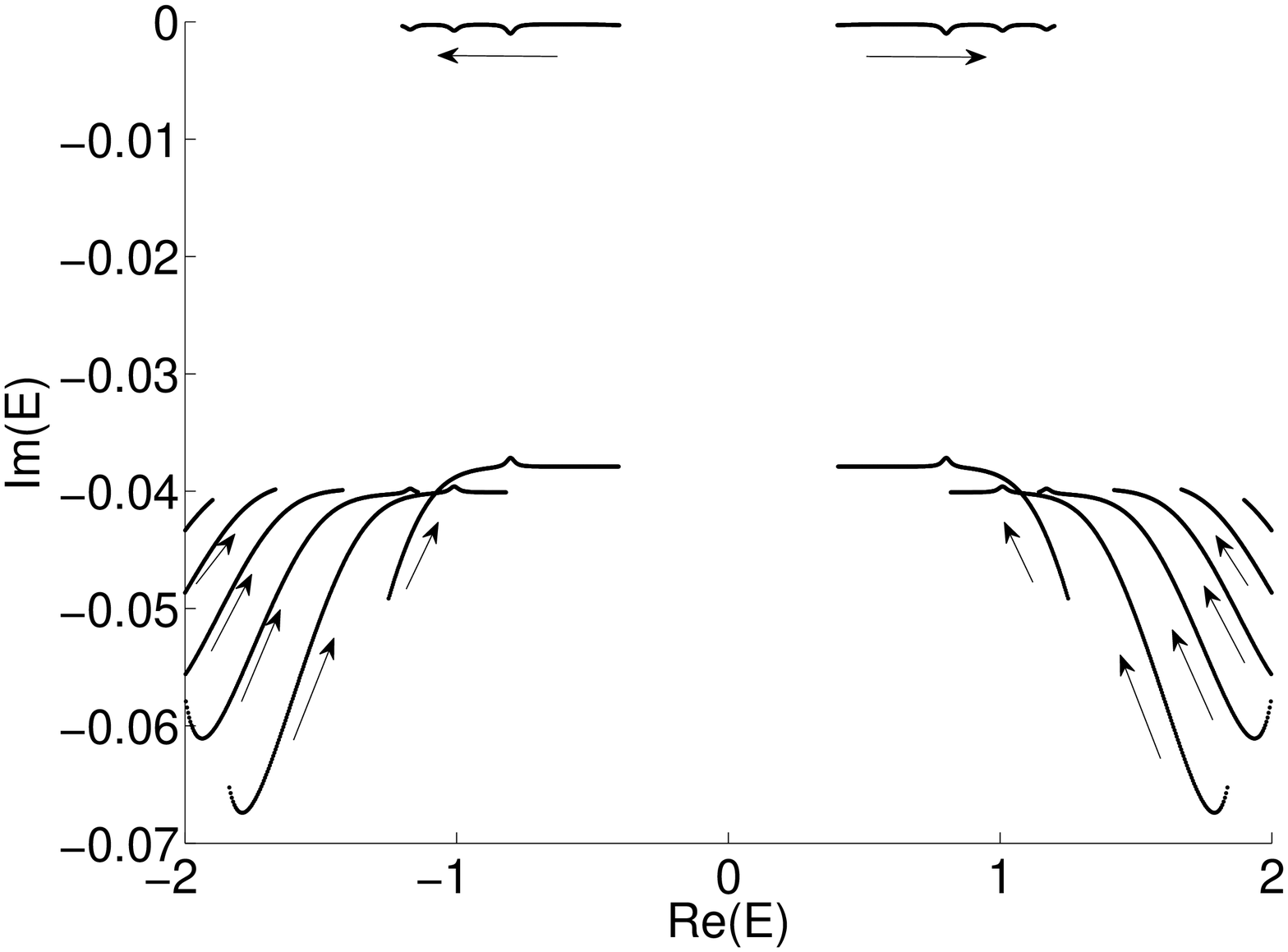}}
\subfloat[]{\includegraphics[width=3.0in]{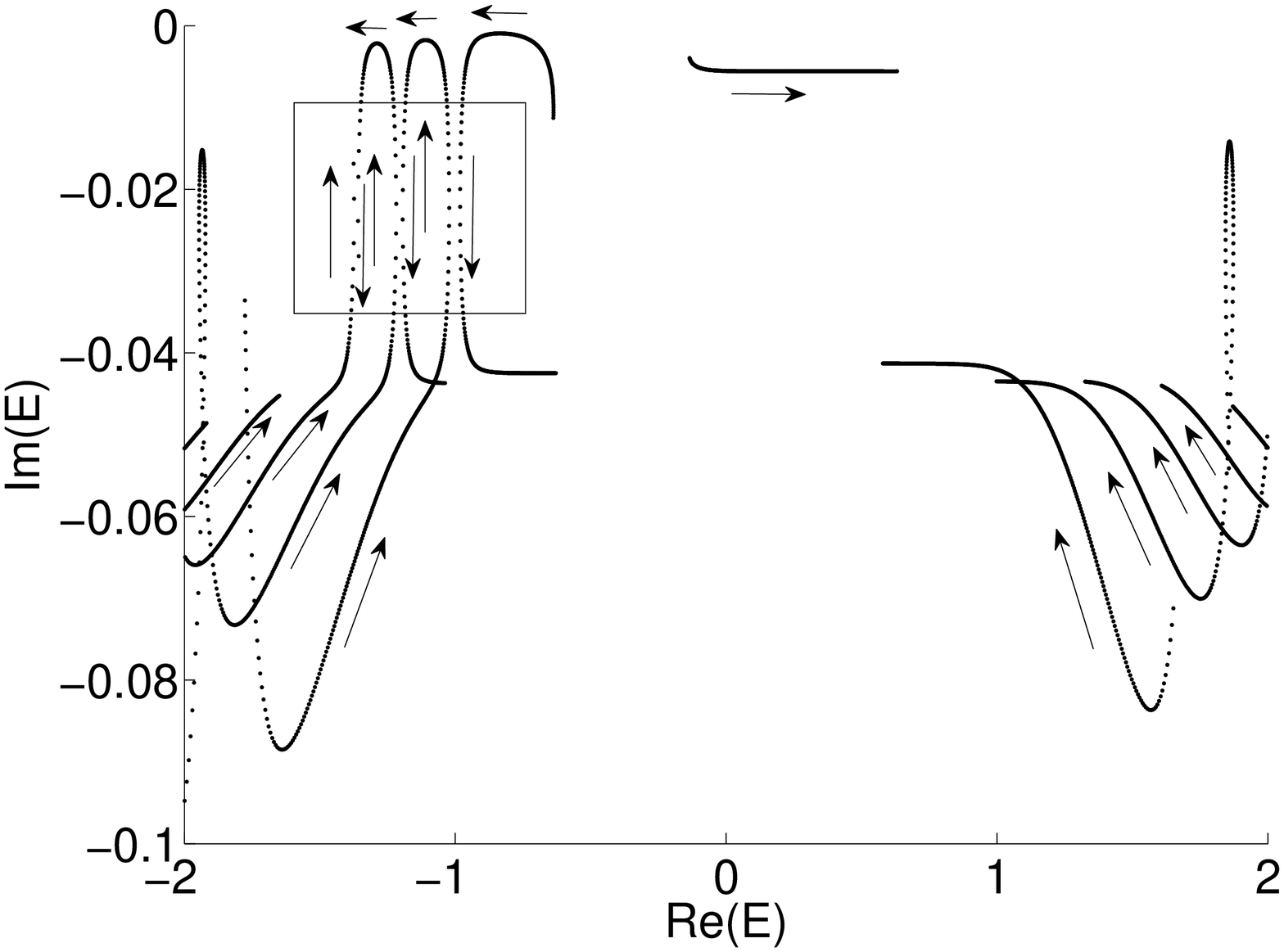}}
\caption{Position of the poles in the complex energy plane for different values of $R$. In  (a), we have $g=2.0$ while for (b), we have $g=3.0$.  The arrows point towards the direction of increasing interatomic distance. The rectangle surrounds the region where the avoided crossings occur.}
\label{fig:pole_position_imvsre}
\end{figure}

However, a careful analysis of the position of poles in the complex energy plane reveals that it is otherwise. For a given value of $g$ and as $R$ increases, there are two possibilities: either the ground state resonance goes below $-mc^{2}$ monotonically or either it induces a series of avoided crossings (the crossing of resonances was discussed in \cite{Bulla1988359} without electric field and in \cite{0305-4470-36-8-311} for the non-relativistic case). An example of this is shown in Figure \ref{fig:pole_position} and \ref{fig:pole_position_imvsre} for two values of potential wells strength ($g=2.0$ and $g=3.0$) and where the the electric field strength is fixed to $F=0.2$ while the interatomic distance is varied by increment of $0.01$. The position of poles in the complex plane is obtained by solving numerically the equation 
\begin{eqnarray}
&&\left[iG_{u}^{+} U^{+}_{1}[a,y(R)]  - G_{u}^{-} U^{-}_{1}[a,y(R)] \right]  \left[iH_{v}^{-} U^{+}_{2}[a,y(-R)]  - H_{v}^{+} U^{-}_{2}[a,y(-R)] \right] \nonumber \\
&-& \left[ iH_{u}^{-} U^{+}_{2}[a,y(-R)]  - H_{u}^{+} U^{-}_{2}[a,y(-R)]  \right]  \left[ iG_{v}^{+} U^{+}_{1}[a,y(R)]  - G_{v}^{-} U^{-}_{1}[a,y(R)]  \right] \nonumber \\
&& = 0,
\end{eqnarray}
which corresponds to the denominator of the spectral density. The figures show how the position of the poles varies as a function of $R$ in the complex energy plane. As seen in these pictures, for $g=3.0$, the first avoided crossing occurs when the real part of the bound state resonance approaches $-mc^{2}$. At that point, its imaginary part changes rapidly to take the value that the nearby resonance had before the avoided crossing. In some sense, the two resonances exchange their positions in the complex plane, which is why they were indistinguishable in the peak structure behavior of the spectral density. For $g=2.0$, there is no avoided crossings: the ground state resonance keeps going to lower energy as the interatomic distance is increased. 

The physical implications of this phenomena (such as whether the system becomes overcritical \cite{Greiner:1985} when the ground state resonance reaches $\Re(E^{F}_{\rm ground}) \lesssim -mc^{2}$), its extension to a real 3-D system and a systematic study of the poles behavior for all parameters will be presented elsewhere. 

\section{Conclusion}

In this work, the spectral density for the 1-D Dirac equation was evaluated for three different cases using the WTK method. The free case was treated as a consistent test for the other systems studied. The main new feature in comparison to the non-relativistic case was the appearance of the negative energy continuum, as expected. The double delta potential wells potential was then treated. By varying the parameters of the model, it was possible to investigate the behavior of the symmetric and anti-symmetric states which were characterized by their eigenvalues positioned in the mass gap (although for sufficiently strong or weak potential wells, one of the state disappeared in the continuum). Finally, the last case considered was the same system immersed in an electric field. This allowed to investigate the appearance of Stark resonances and to characterize their behavior. It was shown that they behave in a similar fashion to the non-relativistic case, except when $R$ or $F$ is large enough in which case new relativistic effects were observed.


The main new physical feature of our analysis is the fact that a resonance having the properties of the ground state resonance can move into the negative energy sea (of the free operator) for a sufficiently large interatomic distance and electric field strength. As the interatomic distance is increased, this happens either by simply moving in this energy region or through a sequence of avoided crossings. This phenomenon is clearly absent from the non-relativistic treatment. This suggests that the system becomes overcritical \cite{Greiner:1985} in that case and thus, can produce particle-antiparticle pairs. Strictly speaking, as soon as the electric field is turned on, electron-positron pairs start to be produced from the Schwinger mechanism. When there are no potential wells, the probability of producing a pair is very small and so is the number of pairs produced, unless the field reaches a certain large value. Our analysis seems to imply that the presence of the diatomic molecule would allow decreasing the threshold for pair production by using the Stark effect. However, our work considered only the single particle Dirac equation: a definite conclusion can only be reached by a Quantum Field Theory treatment, which will be the topic of a future publication. Finally, one should also note that only a few qualitative features of the system were presented in this work. An exhaustive study of the resonance positions in the whole parameter space will be the topic of future investigations. 

From the mathematical point of view, our work combined two main mathematical theories: the WTK method and the Colombeau's generalized function theory. The former is well-known and has been used to analyze many physically relevant equations and to obtain analytical solutions, as done in our analysis. To the best of our knowledge, it is the first time the latter is used to treat point interactions in the Dirac equation. Although the final result obtained is not completely new, we hope our analysis puts it on a more rigorous ground. For instance, the usual way of dealing with the electrostatic point interaction is to integrate the Dirac equation around the point interaction, that is on $[-\epsilon,\epsilon]$ (for $V(x)=-g\delta(x)$) and then, make the following assumption \cite{0022-3719-5-8-006,springerlink:10.1007/BF00750071}:
\begin{eqnarray}
\label{eq:assumption}
\int_{0^{-}}^{0^{+}}dx \psi(x) \delta(x) = \frac{1}{2} \left[ \psi(0^{+}) + \psi(0^{-}) \right].
\end{eqnarray}
Recall that $\psi$ has a jump discontinuity at $x=0$ and thus, the latter implies certain assumptions on the distribution representative of the potential and wave function. In our analysis, the Dirac equation is interpreted in the sense of Colombeau's distributions. We then use a symmetry argument and the general properties of self-adjoint extension of point interaction to fix the ambiguity in the product of distributions, allowing us to ``derive'' a relation similar to \eqref{eq:assumption} and to obtain boundary conditions for a self-adjoint operator consistent with an electrostatic potential.

\appendix

%

\section{Colombeau's generalized theory for the treatment of point interactions in the 1-D Dirac equation}
\label{sec:app_bound}

The Dirac equation is given by
\begin{equation}
\label{eq:dirac}
 E\psi(x) = \left[ ic \sigma_{y} \partial_{x} - \sigma_{z} mc^{2} + V(x)  \right] \psi(x).
\end{equation}
For the sake of this argument, we consider only one delta distribution interaction given by
\begin{equation}
 V(x) = -g \delta(x).
\end{equation}
The generalization to two or more delta interactions is straightforward. The goal is to find the boundary conditions for the wave function at $x=0$. We do so by working in Colombeau's algebra which non-canonically embeds the set of Schwartz' distributions, allowing the multiplication of distributions, \cite{Colombeau198396,Colombeau1990,colombeau:315,Colombeau:1985}. First, it was argued in \cite{calkin:737} that contrary to the Schr\"odinger equation, where the wave function is continuous (its derivative has a jump discontinuity), the wave function computed with the Dirac equation has a jump discontinuity. Therefore, when considering the term $V(x) \psi(x)$, we are dealing with a product of distributions. 


The first step to find the boundary conditions is to rewrite  (\ref{eq:dirac}) componentwise. This yields  
\begin{eqnarray}
\label{eq:dirac_exp1}
 E\psi_{1}(x) &=& c\partial_{x} \psi_{2}(x) - mc^{2} \psi_{1}(x) + V(x) \psi_{1}(x),\\ 
\label{eq:dirac_exp2}
 E\psi_{2}(x) &=& -c\partial_{x} \psi_{1}(x) + mc^{2} \psi_{2}(x) + V(x) \psi_{2}(x),
\end{eqnarray}
which are true in the sense of distributions. Then, we are seeking discontinuous solutions of the form \cite{colombeau:315}
\begin{eqnarray}
 \label{eq:sol1_dis}
\psi_{1}(x) &=& \Delta\psi_{1}(x) H(x) + \psi_{1}^{-}(x) \;\;\mbox{where}\;\; \Delta\psi_{1}(x) = \psi_{1}^{+}(x) - \psi_{1}^{-}(x) ,\\
\label{eq:sol2_dis}
\psi_{2}(x) &=& \Delta\psi_{2}(x) K(x) + \psi_{2}^{-}(x) \;\;\mbox{where}\;\; \Delta\psi_{2}(x) = \psi_{2}^{+}(x) - \psi_{2}^{-}(x) ,
\end{eqnarray}
where $\psi_{1,2}^{\pm}$ are smooth continuous functions defined in $\R$ and solution of the free ($g=0$) Dirac equation, $H$ and $K$ are Heaviside generalized functions. \textit{A priori}, the latter are equal in the weak sense only ($K \approx H$): that is there exists $(K_{\epsilon})_{0<\epsilon<1}$, $(H_{\epsilon})_{0<\epsilon<1}$, two distinct representatives respectively of $K$ and $H$, then for all $\phi \in \mathcal{D}(\R)$ (set of $C^{\infty}(\R)$-functions with compact support):
\begin{eqnarray*}
\lim_{\epsilon \rightarrow 0}\int_{\R}K_{\epsilon}(x)\phi(x)dx = K(\phi) = \int_{\R^+}\phi(x) = \lim_{\epsilon \rightarrow 0}\int_{\R}H_{\epsilon}(x)\phi(x)dx = H(\phi)
\end{eqnarray*}
such that $\forall \phi \in \mathcal D(\R)$ :
\begin{eqnarray*}
\lim_{\epsilon \rightarrow 0}\int_{\R} \big(K_{\epsilon}(x)-H_{\epsilon}(x)\big)\phi(x)dx = 0.
\end{eqnarray*}
In the strong sense, these last equalities would require $K_{\epsilon}(x)=H_{\epsilon}(x)$ (see \cite{colombeau:315}, for details on these definitions).
In the strong sense \cite{colombeau:315}, we assume that $H\neq K$. We will show \textit{a posteriori} that they are also equal in the strong sense. Substituting  (\ref{eq:sol1_dis}) and (\ref{eq:sol2_dis}) in (\ref{eq:dirac_exp1}) and (\ref{eq:dirac_exp2}), we get, in Colombeau's distributional sense
\begin{eqnarray}
\label{DD}
A[\psi,E] &:=&  -E\Delta\psi_{1}(x) H(x) +c [\partial_{x}\Delta\psi_{2}(x)] K(x) \nonumber \\
&&    +c\Delta\psi_{2}(x) K'(x)  -  mc^{2} \Delta\psi_{1}(x) H(x)  \nonumber \\
&& -g L'(x)\Delta\psi_{1}(x) H(x) -g L'(x) \psi_{1}^{-}(x) =0,\\
\label{DDD}
B[\psi,E] &:=&  -E\Delta\psi_{2}(x) K(x)   -c [\partial_{x}\Delta\psi_{1}(x)] H(x) \nonumber \\
&&-c\Delta\psi_{1}(x) H'(x)  +  mc^{2}\Delta\psi_{2}(x) K(x)  \nonumber \\
&& -g L'(x)\Delta\psi_{2}(x) K(x) -g L'(x) \psi_{2}^{-}(x) =0,
\end{eqnarray}
where $L$ is also a Heaviside function and $H'=\partial_{x}H,K'=\partial_{x}K,L'=\partial_{x}L$ are delta functions. The fact that $\psi_{1,2}^{\pm}$ are solutions of the free Dirac equation was also used to cancel some terms. Then, to find the boundary conditions, we apply $\phi \in \mathcal{D}(\R)$ to \eqref{DD} and \eqref{DDD}:
\begin{eqnarray}
\tilde{A}_{\epsilon}[\psi,E] &:=&  -E\int_{\R}\Delta\psi_{1}(x) H_{\epsilon}(x)\phi(x)dx   +c \int_{\R}[\partial_{x}\Delta\psi_{2}(x)] K_{\epsilon}(x)\phi(x)dx \nonumber \\
&&   +c\int_{\R}\Delta\psi_{2}(x) K_{\epsilon}'(x)\phi(x)dx -  mc^{2} \int_{\R}\Delta\psi_{1}(x) H_{\epsilon}(x)\phi(x)dx \nonumber \\
&&  -g \int_{\R}L_{\epsilon}'(x)\Delta\psi_{1}(x) H_{\epsilon}(x)\phi(x)dx -g \int_{\R}L_{\epsilon}'(x) \psi_{1}^{-}(x)\phi(x)dx \nonumber \\ &=&0,\\
 \tilde{B}_{\epsilon}[\psi,E] &:=& -E\int_{\R}\Delta\psi_{2}(x) K_{\epsilon}(x)\phi(x)dx  -c \int_{\R}[\partial_{x}\Delta\psi_{1}(x)] H_{\epsilon}(x)\phi(x)dx \nonumber \\
&&   -c\int_{\R}\Delta\psi_{1}(x) H_{\epsilon}'(x)\phi(x)dx +  mc^{2}\int_{\R}\Delta\psi_{2}(x) K_{\epsilon}(x)\phi(x)dx \nonumber \\
&&  -g \int_{\R}L_{\epsilon}'(x)\Delta\psi_{2}(x) K_{\epsilon}(x)\phi(x)dx -g \int_{\R}L_{\epsilon}'(x) \psi_{2}^{-}(x)\phi(x)dx \nonumber \\ &=&0.
\end{eqnarray}
%
%
%
%
Performing the integration and taking the limit $\epsilon \rightarrow 0$, we get
\begin{eqnarray}
\label{eq:bound_gen11}
 \phi(0)\left[ c\psi_{2}^{+}(0)-c\psi_{2}^{-}(0) -g a \psi_{1}^{+}(0)   -g(1-a) \psi_{1}^{-}(0) \right] \nonumber \\
- \int_{\R^{+}} \left[ (E+mc^2) \Delta\psi_{1}(x) -c\partial_{x}\Delta\psi_{2}(x) \right] \phi(x)dx  =0, \\
 \nonumber \\
\label{eq:bound_gen12}
 \phi(0)\left[ -c\psi_{1}^{+}(0) + c \psi_{1}^{-}(0)   -g b \psi_{2}^{+}(0) -g (1-b) \psi_{2}^{-}(0) \right]  \nonumber \\
- \int_{\R^{+}} \left[ (E-mc^2) \Delta\psi_{2}(x) +c\partial_{x}\Delta\psi_{1}(x) \right] \phi(x)dx  =0  , 
\end{eqnarray}
where we used the fact that $\psi^{\pm}$ are smooth and where $a,b$ are real constants. These constants are obtained from Colombeau's theory which states that for all $\phi \in \mathcal{D}(\R)$
\begin{eqnarray}
\lim_{\epsilon \rightarrow 0}\int_{\R} L_{\epsilon}'(x)\phi(x) H_{\epsilon}(x) dx &=& a \int_{\R} L_{\epsilon}'(x)\phi(x)  dx = a \phi(0) ,\\ 
\lim_{\epsilon \rightarrow 0}\int_{\R}  L_{\epsilon}'(x)\phi(x) K_{\epsilon}(x) dx &=& b \int_{\R} L_{\epsilon}'(x)\phi(x)  dx = b \phi(0) ,
\end{eqnarray}
for real constants $a,b$ which value depend on the local structure of distributions. Now, we can get rid of the integral terms in \eqref{eq:bound_gen11}, \eqref{eq:bound_gen12} by using again the fact that $\psi_{1,2}^{\pm}$ are solutions of the free Dirac equation and that the equality is satisfied for all $\phi$. We get:
\begin{eqnarray}
\label{eq:bound_gen1}
 c\psi_{2}^{+}(0)-c\psi_{2}^{-}(0) -g a \psi_{1}^{+}(0)   -g(1-a) \psi_{1}^{-}(0)  =0,\\
\label{eq:bound_gen2}
 -c\psi_{1}^{+}(0) + c \psi_{1}^{-}(0)   -g b \psi_{2}^{+}(0) -g (1-b) \psi_{2}^{-}(0)  =0.
\end{eqnarray}
Thus, we have the boundary conditions or jumps conditions on the wave function since we have $\lim_{\epsilon \rightarrow 0} \psi( \pm \epsilon) = \psi^{\pm}(0)$ from \eqref{eq:sol1_dis} and \eqref{eq:sol2_dis}. They are ambiguous because the value of $a,b$ is unknown: they should be fixed by some other physical or mathematical arguments. This is done in the following.


There are many ways to implement point interactions in the 1-D Dirac equation and each corresponds to a different self-adjoint extension of the Dirac operator. It was shown in \cite{Benvegnu1994} that the most general self-adjoint extension is given by a four-parameter family $(\alpha,\delta,\gamma,\beta)$: for each quadruple corresponds a different physical system. This is believed to be the main reason for the ambiguities arising in the definition of the boundary conditions. More precisely, the operator implementing point interactions is given by the free Dirac operator $H$ on the domain \cite{Benvegnu1994}
\begin{eqnarray}
 \mathcal{D}(H) = \left\{ \psi \in \left\{ W^{2,1}(\R \backslash \{0\}) \cap AC_{\rm loc} (\R \backslash \{0\})  \right\} \otimes \mathbb{C}^{2} | \psi(0^{+}) = \Lambda \psi(0^{-})  \right\}
\end{eqnarray}
where $W^{2,1}$ is the Sobolev space, $AC_{\rm loc}$ is the set of absolute continuous function and $\Lambda$ is a $2 \times 2$ matrix implementing the following transformation:
\begin{eqnarray}
\label{eq:gen_bound}
 \psi_{1}(0^{+}) = \omega \left[ \delta \psi_{1}(0^{-}) + \gamma \psi_{2}(0^{-}) \right],\\
\label{eq:gen_bound2}
\psi_{2}(0^{+}) = \omega \left[ \beta \psi_{1}(0^{-}) + \alpha \psi_{2}(0^{-}) \right] ,
\end{eqnarray}
where $\alpha,\delta,\gamma,\beta \in \mathbb{R}$, $\alpha \delta - \gamma \beta = 1$, $\omega \in \mathbb{C}$ and $|\omega|=1$. This choice of parameters guarantees that the Dirac operator is self-adjoint and reduces to the usual delta interaction of the Schr\"odinger equation in the non-relativistic limit. Therefore, the jump condition derived previously in \eqref{eq:bound_gen1} and \eqref{eq:bound_gen1} has to be consistent with these conditions to keep the self-adjointness of the operator. Moreover, it should also fulfill an important ``physical'' condition: the charge conjugation invariance. The latter guarantees that the particle-antiparticle duality is well-defined. Both of these requirements allow to fix the value of $a,b$ uniquely, as shown in the following.

First, we show that $K = H$ in the strong sense by demonstrating that $a=b$ using the charge conjugation invariance. It can be easily found from the time-dependent Dirac equation in our representation that the wave function transforms as $\mathcal{C}\psi(x) = \sigma_{x}\psi^{*}(x)$ under the charge conjugation operator $\mathcal{C}$. Recall that the potential then transforms as $g \rightarrow -g$. Then,  (\ref{eq:bound_gen1}) and (\ref{eq:bound_gen2}) are written in the same form as  (\ref{eq:gen_bound}) and (\ref{eq:gen_bound2}), which yields
\begin{eqnarray}
 \psi_{1}(0^{+}) &=& \left[ 1+\frac{abg^{2}}{c^{2}} \right]^{-1} \left[ \left( 1- \frac{b(1-a)g^{2}}{c^{2}} \right) \psi_{1}(0^{-}) - \frac{g}{c} \psi_{2}(0^{-})  \right] ,\\
\psi_{2}(0^{+}) &=& \left[ 1+\frac{abg^{2}}{c^{2}} \right]^{-1} \left[ \left( 1- \frac{a(1-b)g^{2}}{c^{2}} \right) \psi_{2}(0^{-}) + \frac{g}{c} \psi_{1}(0^{-})  \right],
\end{eqnarray}
and
\begin{eqnarray}
 \delta &=& \left[1 + \frac{g^{2}ab}{c^{2}} \right]^{-1} \left[1- \frac{g^{2}b(1-a)}{c^{2}} \right], \\
\alpha &=& \left[1 + \frac{g^{2}ab}{c^{2}} \right]^{-1} \left[1- \frac{g^{2}a(1-b)}{c^{2}} \right],  \\
 \gamma &=& -\left[1 + \frac{g^{2}ab}{c^{2}} \right]^{-1} \frac{g}{c} = -\beta, \\ 
 \omega &=& 1.
\end{eqnarray} 
We require that the operator defined on the domain $\mathcal{D}(H)$ is invariant under the charge conjugation symmetry, which implies that the matrices $\Lambda^{\mathcal{C}} = \mathcal{C} \Lambda \mathcal{C}^{-1}$ should be the same as $\Lambda$ \cite{springerlink:10.1023/A:1007493325970} such that the following equality holds:
\begin{eqnarray}
 \Lambda := \omega 
\begin{bmatrix}
 \delta & \gamma\\
\beta & \alpha
\end{bmatrix}
= \Lambda^{\mathcal{C}} :=\sigma_{x} \omega^{*}
\begin{bmatrix}
 \delta & -\gamma\\
-\beta & \alpha
\end{bmatrix}
\sigma_{x}.
\end{eqnarray}
Note that the sign of $\beta,\gamma$ in $\Lambda^{\mathcal{C}}$ changes because $g \rightarrow -g$. The last equation implies that
\begin{eqnarray}
\omega 
\begin{bmatrix}
 \delta & \gamma\\
\beta & \alpha
\end{bmatrix}
= \omega^{*}
\begin{bmatrix}
 \alpha & -\beta\\
-\gamma & \delta
\end{bmatrix}
\end{eqnarray}
which is fulfilled if and only if $a=b$.

From this last argument, we obtain
\begin{eqnarray}
 \delta &=& \left[1 + \frac{g^{2}a^{2}}{c^{2}} \right]^{-1} \left[1- \frac{g^{2}a(1-a)}{c^{2}} \right] = \alpha ,\\
 \gamma &=& -\left[1 + \frac{g^{2}a^{2}}{c^{2}} \right]^{-1} \frac{g}{c} = -\beta ,\\ 
 \omega &=& 1,
\end{eqnarray}
which is still ambiguous because the value of $a$ is unknown. However, recall that the self-adjoint extension is defined such that the condition $\alpha \delta - \gamma \beta = 1$ is fulfilled. Substituting the last three equations in this condition, we find that $a$ is the solution of the cubic equation
\begin{eqnarray}
 a^{3} -\frac{1}{2}a^{2} + \frac{c^{2}}{g^{2}} - \frac{c^{2}}{2g^{2}} = 0 ,
\end{eqnarray}
which has two complex solutions and one real solution given by $\frac{1}{2},i\frac{g}{c},-i\frac{g}{c}$. According to Colombeau's theory, the constant value should be a real number so we select $a=\frac{1}{2}$. This fixes the value of $a,b$ uniquely and unambiguously.

To summarize, we have the following boundary conditions:
\begin{eqnarray}
\label{eq:bound_cond1}
 \psi_{1}(0^{+}) &=& \left[ 1+\frac{g^{2}}{4c^{2}} \right]^{-1} \left[ \left( 1- \frac{g^{2}}{4c^{2}} \right) \psi_{1}(0^{-}) - \frac{g}{c} \psi_{2}(0^{-})  \right] ,\\
\label{eq:bound_cond2}
\psi_{2}(0^{+}) &=& \left[ 1+\frac{g^{2}}{4c^{2}} \right]^{-1} \left[ \left( 1- \frac{g^{2}}{4c^{2}} \right) \psi_{2}(0^{-}) + \frac{g}{c} \psi_{1}(0^{-})  \right],
\end{eqnarray}
and also, it is convenient to invert these to get
\begin{eqnarray}
\label{eq:bound_cond3}
 \psi_{1}(0^{-}) &=& \left[ 1+\frac{g^{2}}{4c^{2}} \right]^{-1} \left[ \left( 1- \frac{g^{2}}{4c^{2}} \right) \psi_{1}(0^{+}) + \frac{g}{c} \psi_{2}(0^{+})  \right], \\
\label{eq:bound_cond4}
\psi_{2}(0^{-}) &=& \left[ 1+\frac{g^{2}}{4c^{2}} \right]^{-1} \left[ \left( 1- \frac{g^{2}}{4c^{2}} \right) \psi_{2}(0^{+}) - \frac{g}{c} \psi_{1}(0^{+})  \right],
\end{eqnarray}
These boundary conditions are consistent with the ones found in \cite{0022-3719-5-8-006,RhondaJ1997425} for static electric potential which were derived by using other but similar techniques.

We conclude this Appendix by a remark about the definition of the product of distributions. In the framework of nonlinear hyperbolic systems, Dal Maso, LeFloch and Murat have proposed a definition of such products (typically $H \times \delta$) as a bounded Borel measure dependent of a path $\Phi$ (which has several properties \cite{DLM}) as follows. Assuming that $\psi$ is regular expected at $x_0$ where it has a jump, and assuming $F$ to be a regular function, then the product $F(\psi)\psi'$ is defined as the following measure. If $\psi$ is continuous on a Borel set $B$, then
\begin{eqnarray*}
[F(\psi)\psi']_{\Phi}(B) = \int_B F(\psi)\psi'(x) dx
\end{eqnarray*}
At $x_0$, where $\psi$ is discontinuous
\begin{eqnarray*}
[F(\psi)\psi']_{\Phi}({x_0}) = \int_0^1 F\big(\Phi(s;\psi(x_0^-),\psi(x_0^+)\big)\cfrac{\partial \Phi}{\partial s}(s;\psi(x_0^-),\psi(x_0^+))ds
\end{eqnarray*}
where by definition $\Phi\big(0;\psi(x_0^-),\psi(x_0^+)\big)=\psi(x_0^-)$ and 
$\Phi\big(1;\psi(x_0^-),\psi(x_0^+)\big)=\psi(x_0^+)$. If $\psi(x):= \psi_0 + H(x-x_0)(\psi_1-\psi_0)$
\begin{eqnarray*}
[F(\psi)\psi']_{\Phi} =\int_0^1 F\big(\Phi(s;\psi_0,\psi_1)\big)\cfrac{\partial \Phi}{\partial s}(s;\psi_0,\psi_1)ds  \times \delta_{x_0}
\end{eqnarray*}
Now in the case, where $F$ is the identity, we get
\begin{eqnarray*}
[\psi\psi']_{\Phi} =\int_0^1 \Phi(s;\psi_0,\psi_1)\cfrac{\partial \Phi}{\partial s}(s;\psi_0,\psi_1)ds  \times \delta_{x_0} = \cfrac{1}{2}\delta_{x_0}
\end{eqnarray*}
That is: ``$H\delta = \delta /2$'', which is the same result obtained from Colombeau's theory. Details can be found in \cite{DLM}, \cite{LF}.

\begin{acknowledgments}
The authors would like to thank Nico Temme for his help with parabolic cylinder functions. We also thank Fritz Gesztesy for useful discussions. 
\end{acknowledgments}

\bibliography{bibliography}

\end{document}